%% file: arXiv_v4.tex
\documentclass[sigconf,balance=false,natbib=false,acmthm=false]{acmart}
\AtBeginDocument{%
  \providecommand\BibTeX{{%
    \normalfont B\kern-0.5em{\scshape i\kern-0.25em b}\kern-0.8em\TeX}}}

\usepackage{latexsym}
\usepackage{amsmath,mathtools}
\usepackage{amsfonts}
\usepackage{balance}
\usepackage{cite}
\usepackage[linesnumbered,lined,ruled,noend,vlined,boxed]{algorithm2e}
\usepackage{cleveref}
\usepackage{tabularx,environ}
\usepackage{comment}
\usepackage{url}
\usepackage{lscape}
\usepackage{lineno}
\usepackage[normalem]{ulem}
\usepackage{arydshln}
\usepackage{mathtools}
\usepackage{multicol}

\usepackage{multirow}
\usepackage{booktabs}
\usepackage{caption} 
\usepackage{color}
\definecolor{darkred}{rgb}{0.8,0,0}
\newcommand{\revised}[1]{{\color{black} #1}}

\input{macro}
\usepackage[appendix=inline]{apxproof} 
\newtheoremrep{theorem}{Theorem}
\newtheoremrep{lemma}[theorem]{Lemma}
\crefname{lemmarep}{lemma}{lemmas}
\newtheorem{proposition}[theorem]{Proposition}
\newtheorem{definition}[theorem]{Definition}

\makeatletter
\newtheorem*{restate@theorem}{\restate@title}
\newcommand{\newrestatetheorem}[2]{%
\newenvironment{restate#1}[1]{%
 \def\restate@title{#2 \ref{##1}}%
 \begin{restate@theorem}}%
 {\end{restate@theorem}}}
\makeatother
\newrestatetheorem{theorem}{Theorem}

\begin{document}
\fancyhead{} 

\title{Linear-Delay Enumeration for Minimal Steiner Problems}



\author{Yasuaki Kobayashi}
\affiliation{%
  \institution{Hokkaido University}
  \streetaddress{}
  \city {Hokkaido}
  \country{Japan}}
\email{koba@ist.hokudai.ac.jp}

\author{Kazuhiro Kurita}
\affiliation{%
  \institution{National Institute of Informatics}
  \streetaddress{Hitotsubashi, Chiyoda-ku, Tokyo}
  \city{Tokyo}
  \country{Japan}}
\email{kurita@nii.ac.jp}

\author{Kunihiro Wasa}
\affiliation{%
  \institution{Toyohashi University of Technology}
  \streetaddress{Hibarigaoka, Tenpakucho, Toyohashi, Aichi}
  \city{Aichi}
  \country{Japan}}
\email{wasa@cs.tut.ac.jp}

\renewcommand{\shortauthors}{Y. Kobayashi, K. Kurita, and K. Wasa}
\begin{abstract}
    Kimelfeld and Sagiv [Kimelfeld and Sagiv, PODS 2006], [Kimelfeld and Sagiv, Inf. Syst. 2008] pointed out \revised{that} the problem of enumerating $K$-fragments is of great importance in a keyword search on data graphs.
    In a graph-theoretic term, the problem corresponds to enumerating minimal Steiner trees in (directed) graphs.
    In this paper, we propose a linear-delay and polynomial-space algorithm for enumerating all minimal Steiner trees, improving on a previous result in [Kimelfeld and Sagiv, Inf. Syst. 2008].
    Our enumeration algorithm can be extended to other Steiner problems, such as minimal Steiner forests, minimal terminal Steiner trees, and minimal directed Steiner trees.
    As another variant of the minimal Steiner tree enumeration problem, we study the problem of enumerating minimal induced Steiner subgraphs. 
    We propose a polynomial-delay and exponential-space enumeration algorithm of minimal induced Steiner subgraphs on claw-free graphs.
    Contrary to these tractable results, we show that the problem of enumerating minimal group Steiner trees is at least as hard as the minimal transversal enumeration problem on hypergraphs.
\end{abstract}



\maketitle

\section{Introduction}

Kimelfeld and Sagiv~\cite{Kimelfeld:Efficiently:2008} observed that enumerating \emph{$K$-fragments} for a set of keywords $K$ in data graphs is a core component in several keyword search systems. 
A data graph is a graph that consists of two types of nodes: structural nodes and keyword nodes, and each keyword node corresponds to some keyword in $K$.
\revised{A $K$-fragment is a subtree in a data graph that contains all keyword nodes for $K$ and no proper subtree that contains them. } 
There are several types of $K$-fragments, \emph{undirected $K$-fragments}, \emph{strong $K$-fragments}, and \emph{directed $K$-fragments}.
In a graph-theoretic term, they are equivalent to \emph{Steiner trees}, \emph{terminal Steiner trees}, and \emph{directed Steiner trees}, respectively.

Given an undirected graph $G = (V, E)$ and a subset of vertices $W \subseteq V$, called \emph{terminals},   
the \emph{Steiner tree problem} asks to find a \revised{Steiner tree} of $(G, W)$ that has a minimum number of edges.
Here, a \revised{\emph{Steiner tree} of $(G, W)$ is a subtree of $G$ that contains all terminals}.
The Steiner tree problem is a classical combinatorial optimization problem and has arisen in several areas~\cite{DBLP:journals/ipl/LinX02,Brazil2001,DBLP:journals/mp/GrotschelMW97,Kimelfeld:Efficiently:2008,DBLP:conf/pods/KimelfeldS06}.
This problem is shown to be \NP-hard in Karp's seminal work~\cite{Karp:Reducibility:1972} and has been studied from several perspectives, such as approximation algorithms~\cite{Byrka:Steiner:2013,Goemans:Matroids:2012}, parameterized algorithms~\cite{DBLP:journals/networks/DreyfusW71}, and algorithms in practice~\cite{Beyer:Strong:2019,DBLP:conf/aaai/IwataS19,Sasaki:Cost:2021}.
There are also many variants of this problem. 
See~\cite{Hauptmann:compendium:2015}, for a compendium on variants of the Steiner tree problem.

The notion of Steiner trees emerges in investigating the connectivity or reachability of a specified vertex subset in networks. 
Steiner trees can be seen as a generalization of some basic combinatorial structures in graphs, such as $s$-$t$ (shortest) paths and spanning trees. 
In some applications, it is preferable to find \emph{multiple} solutions rather than a single solution.
E.g., the problem of finding $k$ distinct shortest $s$-$t$ paths has been widely studied~\cite{ZIJPP:Catalano:TRPB:2020,Hershberger:Maxel:Suri:ACM:Trans:2007,Eppstein:SIAM:J:COMP:1998} due to various practical applications. 

However, as mentioned before, finding a \emph{minimum} Steiner subgraph is intractable in general.
Motivated by these facts, we focus on enumerating \emph{minimal} Steiner subgraphs. 
We say that a Steiner subgraph $H$ of $(G, W)$ is \emph{minimal} if there is no proper subgraph of $H$ that is a Steiner subgraph of $(G, W)$.
It is easy to see that every minimal Steiner subgraph of $(G, W)$ forms a tree, but some Steiner trees of $(G, W)$ may not be minimal Steiner subgraphs of $(G, W)$. 
In this paper, we address the problem of enumerating minimal Steiner trees of $(G, W)$, which is defined as follows.

\begin{definition}[{\sc Steiner Tree Enumeration}]
    \label{def:ste}
    Given an undirected graph $G = (V, E)$ and a terminal set $W \subseteq V$, the task is to enumerate all the minimal Steiner trees of $(G, W)$.
\end{definition}

\begin{table*}[t]
\centering
    \begin{tabular}{l|c|c|c}
        Enumeration problem & Time & Preprocessing & Space \\\hline
        \textsc{Steiner Tree}~\cite{Kimelfeld:Efficiently:2008} & \revised{$\order{m(\size{T_i} + \size{T_{i-1}})}^\dagger$} & $\order{m\size{T_1}}$ & $\order{nm}$ \\ 
        \hdashline
        \textsc{Steiner Forest}~\cite{DBLP:journals/siamdm/KhachiyanBEGM05} & inc. poly. & poly. & exp. \\  
        \hdashline
        \textsc{Terminal Steiner Tree}~\cite{Kimelfeld:Efficiently:2008} & \revised{$\order{m(\size{T_i} + \size{T_{i-1}})}^\dagger$}  &  $\order{m\size{T_1}}$ & $\order{nm}$ \\  
        \hdashline
        \textsc{Directed Steiner Tree}~\cite{Kimelfeld:Efficiently:2008} &  \revised{$\order{mt(\size{T_i} + \size{T_{i-1}})}^\dagger$} & $\order{mt\size{T_1}}$ & $\order{nm}$ \\  
        \hdashline
        \textsc{Induced Steiner Subgraph} with $t \le 3$~\cite{DBLP:conf/mfcs/ConteGKMUW19} & $\order{mn^3}$ amortized & $\order{mn^3}$ & poly. \\
        \hline
        \textsc{Minimum Steiner Tree}~\cite{DBLP:journals/anor/DouradoOP14} & $\order{n}$ delay & $\order{n^{t-2} + n^2m}$ & $\order{n^{t-2}+n^2m}$ \\ 
        \hdashline
        \textsc{Minimum Induced Steiner Subgraph}~\cite{DBLP:conf/mfcs/ConteGKMUW19} & $\order{m2^{3t^2}}$ amortized & $\order{n^{t-2} + n^2m}$ & $\order{n^{t-2}+n^2m}$ \\
        \hline
        \textsc{Steiner Tree} [This work] & $\order{n + m}$ delay & $\order{n(n + m)}$ & $\order{n^2}$ \\ 
        \hdashline
        \textsc{Steiner Forest} [This work] & $\order{n + m}$ delay & $\order{n(n + m)}$ & $\order{n^2}$ \\
        \hdashline
        \textsc{Terminal Steiner Tree}[This work] & $\order{n + m}$ delay & $\order{n(n + m)}$ & $\order{n^2}$ \\  
        \hdashline
        \textsc{Directed Steiner Tree} [This work] & $\order{n + m}$ delay & $\order{n(n + m)}$ & $\order{n^2}$  \\  
        \hdashline
        \textsc{Induced Steiner Subgraph} on claw-free graphs [This work] & poly. delay& poly. & exp. \\
    \end{tabular}
    \caption{The table summarizes known and our results for {\sc Steiner Tree Enumeration} and related problems. Let $n$, $m$, and $t$ be the number of vertices, edges, and terminals, respectively.
    $|T_i|$ the number of edges in the $i$-th solution. 
    Note that the algorithms for {\sc Minimum Steiner Tree} and {\sc Minimum Induced Steiner Subgraph} enumerate all \emph{minimum} Steiner trees, whereas algorithms for other problems enumerate \emph{minimal} Steiner trees.
    \revised{The running time marked $\dagger$ indicates the delay between the $(i-1)$-th and the $i$-th solutions.}
    }
    \label{tab:related}
\end{table*}

There are several known results on enumeration algorithms related to {\sc Steiner Tree Enumeration} and its variants~\cite{DBLP:conf/pods/KimelfeldS06,DBLP:journals/anor/DouradoOP14,DBLP:journals/siamdm/KhachiyanBEGM05,DBLP:conf/mfcs/ConteGKMUW19,Kimelfeld:Efficiently:2008}. See Table~\ref{tab:related} for details.
We say that a Steiner tree $T$ of $(G, W)$ is a \emph{terminal Steiner tree} of $(G, W)$ if every terminal in $W$ is a leaf of $T$.
For a collection of terminal sets $\mathcal W = \{W_1, \ldots, W_s\}$, where $W_i \subseteq V$, we say that a subgraph of $G$ is a \emph{Steiner forest} of $(G, \mathcal W)$ if for each $1 \le i \le s$, the subgraph has a component that is a Steiner subgraph of $(G, W_i)$.
For a directed graph $D = (V, E)$, a terminal set $W$, and $r \in V \setminus W$, we say that a subgraph of $D$ is a \emph{directed Steiner tree} of $(D, W, r)$ if the subgraph contains directed path from $r$ to each $w \in W$.
Since minimal subgraphs of each variant form trees, forests, and directed trees, these are called terminal Steiner trees, Steiner forests, and directed Steiner trees, respectively.

For enumeration problems, {\em enumeration algorithms} are required to generate all the solutions one by one without duplication.
Since enumeration problems generally have an exponential number of solutions, we measure the complexity of enumeration algorithms in terms of the input size $n$ and the output size $N$ (i.e., the number of solutions).
The \emph{delay} between two consecutive solutions of an enumeration algorithm is the running time interval between them.
The delay of an enumeration algorithm is the worst delay between every pair of consecutive solutions.
Note that, unless stated otherwise, the running time before generating the first solution and after generating the last solution is upper bounded by the delay.
If the delay between $(i-1)$-th solution and $i$-th solution is bounded by $\order{poly(n + i)}$, then we call such an enumeration algorithm an \emph{incremental-polynomial time algorithm}.
An enumeration algorithm is called a \emph{polynomial-delay enumeration algorithm} if the delay is upper bounded by a polynomial in $n$.
Finally, an enumeration algorithm that runs in total $\order{N \cdot poly(n)}$ time is called an \emph{amortized polynomial-time algorithm}.

Kimelfeld and Sagiv developed three \revised{efficient} enumeration algorithms for {\sc Steiner Tree Enumeration} and its variants~\cite{Kimelfeld:Efficiently:2008}.
Their algorithms \revised{for} enumerating minimal Steiner trees, minimal terminal Steiner trees, and minimal directed Steiner trees run in $O(m(|T_i| + |T_{i-1}|))$, $O(m(|T_i| + |T_{i-1}|))$, and $O(mt(|T_i| + |T_{i-1}|))$ \revised{delay between $(i-1)$-th and $i$-th solutions}, respectively.
Here, $m$ is the number of edges in the input graph, $|T_i|$ is the number of edges in the $i$-th solution, and $t$ is the number of terminals, respectively. 
They also proposed efficient algorithms for enumerating Steiner trees and their variants in an ``approximate'' ascending order of their weights~\cite{DBLP:conf/pods/KimelfeldS06}.
Khachiyan et al.~\cite{DBLP:journals/siamdm/KhachiyanBEGM05} studied the problem of enumerating minimal Steiner forests in graphs as a special case of the circuit enumeration problem on matroids and gave an incremental-polynomial time enumeration algorithm for this problem.

\medskip
\noindent
\textsf{\textbf{Our contribution: }} Here, we give our main result for {\sc Steiner Tree Enumeration}. 
\begin{theorem}\label{theo:st:summary}
    There is an $O(n + m)$ delay and $\order{n^2}$
    space\footnote{Whenever we refer to space complexity, we assume \revised{the RAM model, that is, }a single word of memory contains $O(\log n)$ bits.} enumeration algorithm for {\sc Steiner Tree Enumeration} provided that we are allowed to use $O(n(n + m))$ preprocessing time, where $n$ and $m$ are the number of vertices and edges, respectively.
    Moreover, without additional preprocessing time, we can implement the algorithm so that it runs in time $O(n + m)$ amortized time per solution and $O(n + m)$ space.
\end{theorem}
This algorithm can be extended \revised{to} enumerating minimal Steiner forests, minimal terminal Steiner trees, and minimal directed Steiner trees.
Our algorithms for these problems achieve the same running time bounds \revised{as in \Cref{theo:st:summary}. (See \Cref{thm:sfe:delay,thm:tst,thm:dst}.)}

The basic idea of these algorithms is a standard branching technique.
Starting from an arbitrary terminal, we recursively grow a partial Steiner tree $T$ by attaching a path between $T$ and a terminal not contained in $T$.
Since the paths between $T$ and a terminal $w$ can be enumerated in polynomial delay, we immediately have a polynomial-delay and polynomial-space algorithm for {\sc Steiner Tree Enumeration}.
To improve the running time and \revised{space complexity} of this simple algorithm, we carefully design the entire branching strategy and the path enumeration algorithm.
\revised{
To this end, we employ Read and Tarjan's path enumeration algorithm~\cite{Read:Tarjan:Networks:1975} and discuss the delay and space complexity of their algorithm, which is not discussed explicitly in \cite{Read:Tarjan:Networks:1975}. 
As for the delay complexity, we can easily obtain a linear delay bound by applying a standard technique~\cite{Kimelfeld:Efficiently:2008,Uno::2003}.
However, to obtain a linear space bound for {\sc Steiner Tree Enumeration}, we need a nontrivial modification of their algorithm, which will be discussed in \Cref{sec:path:enum}.
}
We also exploit the output queue method due to Uno~\cite{Uno::2003} to improve the delay for {\sc Steiner Tree Enumeration}.

\revised{
    In contrast to these tractable cases, we also show that \revised{two} variants of {\sc Steiner Tree Enumeration} are ``not so easy'': the problem of enumerating internal Steiner trees is \NP-hard, and that of enumerating group Steiner trees is at least as hard as the minimal hypergraph transversal problem (see \Cref{sec:preliminaries,sec:hardness} for their definitions).
    \revised{It should be mentioned that the polynomial-time solvability for the minimal hypergraph transversal enumeration problem with respect to the input size $n$ and output size $N$ (i.e., $(n + N)^{O(1)}$) is one of the most challenging open problems in the field of enumeration algorithms and the best known algorithm runs in quasi-polynomial time $(n+N)^{o(\log (n + N))}$~\cite{Fredman:complexity:1996}.}
}

We also consider the problem of enumerating minimal induced Steiner subgraphs as another variant of {\sc Steiner Tree Enumeration}.
This problem is known to be at least as hard as the minimal transversal enumeration problem on hypergraphs even when the input is restricted to split graphs~\cite{DBLP:conf/mfcs/ConteGKMUW19}. 
They also showed that if the number of terminals is at most three, one can solve this problem in $\order{mn^3}$ amortized time per solution \revised{and polynomial space}. 
In this paper, we develop a polynomial-delay and exponential-space enumeration algorithm on claw-free graphs with an arbitrary number of terminals. 
Since the class of claw-free graphs is a superclass of line graphs~\cite{BEINEKE1970129} and {\sc Steiner Tree Enumeration} is a special case of the problem of enumerating minimal induced Steiner subgraphs on line graphs, our result non-trivially expands the tractability of Steiner subgraph enumeration.

\section{Preliminaries}
\label{sec:preliminaries}
Let $G = (V, E)$ be an undirected (or directed) graph with vertex set $V$ and edge set $E$.
We also denote by $V(G)$ and $E(G)$ the sets of vertices and edges in $G$, respectively. 
Throughout the paper, we assume that graphs have no self-loops \revised{but may have multiedges}.
For a vertex $v$, we denote by $N_G(v)$ the set of neighbors of $v$ in $G$ and by $\Gamma_G(v)$ the set of incident edges of $v$.
These notations are extended to sets: For $U \subseteq V$, $N_G(U) = \bigcup_{v \in U} N_G(v) \setminus U$ and $\Gamma_G(U) = \bigcup_{v \in U} \Gamma_G(v) \setminus \{\{u, v\} : u, v \in U\}$.
If there is no confusion, we drop the subscript $G$ from these notations.
For an edge $e = \set{u, v}$, the graph obtained from $G$ by contracting $e$ is denoted by $G \slash e$, that is, $G \slash e = (V', E')$, where $V' = (V \setminus \set{u, v}) \cup \set{w}$ and \revised{$E' = (E \setminus (\Gamma_G(u) \cup \Gamma_G(v))) \cup \{\{x, w\} : \{x, y\} \in \Gamma_G(u) \cup \Gamma_G(v) \land ( y = u \lor y = v) \}$}.
\revised{In this notation, the set $E'$ is considered as a multiset and hence} $G \slash e$ may have multiedges.
For a set of edges $F \subseteq E$, the (multi)graph $G \slash F$ is obtained from $G$ by contracting all edges in $F$.

For $U \subseteq V$, $G[U]$ is the subgraph of $G$ induced by $U$, that is, $G[U] = (U, E')$ with $E' = \{\{u, v\} \in E : u, v \in U\}$.
For $F \subseteq E$, we also use $G[F]$ to denote \revised{the subgraph of the form $(V(F), F)$}, where $V(F)$ is the set of end vertices of edges in $F$. 
Let $H$ be a subgraph of $G$.
For $F \subseteq E(G) \setminus E(H)$, we denote by $H + F$ the graph $G[E(H) \cup F]$.
Similarly, for $F \subseteq E(H)$, we denote a subgraph $G[E(H) \setminus F]$ as $H - F$. 
When $F$ consists of a single edge $e$, we simply write $H + e$ and $H - e$ to denote $H + F$ and $H - F$, respectively.
In this paper, we may identify a subgraph with its edges.

Let $X$ and $Y$ be disjoint sets of vertices of $G$.
A path $P$ is called an \emph{$X$-$Y$ path} if an end vertex of $P$ belongs to $X$, the other end belongs to $Y$, and every internal vertex of $P$ does not belong to $X \cup Y$.
If $X$ and/or $Y$ are singletons $u$ and $v$, respectively, we may write $u$-$Y$, $X$-$v$, $u$-$v$ paths to denote an $X$-$Y$ path. 

Let $W \subseteq V$ be a set of terminals.
A subgraph $H$ of $G$ is a \emph{Steiner subgraph of $(G, W)$} if there is a path between every pair of vertices of $W$ in $H$.
Note that every minimal Steiner subgraph forms a tree, called a \emph{minimal Steiner tree}. 
The following characterization is straightforward from the observation that 
every leaf of a minimal Steiner tree belongs to $W$ as otherwise, we can delete it.
\begin{proposition}\label{prop:st:char}
    \revised{A Steiner tree $T$ of $(G, W)$ is minimal if and only if \revised{every leaf of $T$ is a terminal}.}
\end{proposition}

\revised{
Let $\mathcal W = \set{W_i \subseteq V: 1 \le i \le s}$ be a family of terminal sets.
A subgraph of $G$ is called a \emph{Steiner forest} of $(G, \mathcal W)$ if it has no cycles and there is a path between every pair of vertices in $W_i$ for $1 \le i \le s$ in the forest.
We say that a Steiner forest is \emph{minimal} if it has no any Steiner forest as a subgraph.
A subgraph of $G$ is called a \emph{group Steiner tree} of $(G, \mathcal W)$ if it is a tree and there is a path between \emph{some} $u \in W_i$ and \emph{some} $v \in W_j$ for $1 \le i,j \le s$ in the tree. 
We say that a group Steiner tree is \emph{minimal} if it has no any group Steiner tree as a subgraph.
}

Next, we define some terminologies for directed graphs.
Let $D = (V, E)$ be a directed graph.
If $D$ has an edge $(u, v)$, $v$ is called an \emph{out-neighbor} of $u$ and $u$ is called an \emph{in-neighbor} of $v$.
\revised{
    Similarly, $e$ is called an \emph{incoming edge} of $v$ and $e$ is called an \emph{outgoing edge} of $u$.
} 
We say that a vertex $v$ is a \emph{source} (resp. \emph{sink}) if $v$ has no in-neighbors (resp. no out-neighbors) in $D$.
Let $T$ be a subgraph of $D$.
We say that $T$ is a \emph{directed tree rooted at $r$} if $T$ has the unique source $r$ and exactly one directed path from $r$ to each vertex in $T$. 
A vertex $\ell$ in a directed tree is called a \emph{leaf} if $\ell$ has no out-neighbors.
Let $W$ be a set of terminals and let $r \in V \setminus W$. 
We say that a subgraph $H$ is a \emph{directed Steiner subgraph} of $(D, W, r)$ if $H$ has a directed $r$-$w$ path for any terminal $w \in W$.
In particular, a directed Steiner subgraph $T$ of $(D, W, r)$ is a \emph{directed Steiner tree} of $(D, W, r)$ if $T$ forms a directed tree rooted at $r$. 
A directed Steiner subgraph $H$ is \emph{minimal} if no proper subgraph of $H$ is a directed Steiner subgraph of $(D, W, r)$.

In this paper, in addition to {\sc Steiner Tree Enumeration}, we address the following problems.

\begin{definition}[{\sc Steiner Forest Enumeration}]
    \revised{Given an undirected graph $G = (V, E)$ and terminal sets $\mathcal W = \set{W_i \subseteq V: 1 \le i \le s}$, the problem asks to enumerate all the minimal Steiner forest of $(G, \mathcal W)$}.
\end{definition}

\begin{definition}[{\sc Internal Steiner Tree Enumeration}]
\label{def:ISTE}
    \revised{
        Given an undirected graph $G = (V, E)$ and a terminal set $W \subseteq V$, the problem asks to enumerate all the Steiner trees $T$ of $(G, W)$ in which every vertex in $W$ belongs to $T$ as an internal vertex.\footnote{In this definition, solutions are not required to be ``minimal''.}
    }
\end{definition}

\begin{definition}[{\sc Terminal Steiner Tree Enumeration}]
    Given an undirected graph $G = (V, E)$ and a terminal set $W \subseteq V$, the problem asks to enumerate all the minimal Steiner trees $T$ of $(G, W)$ in which every vertex in $W$ belongs to $T$ as a leaf vertex.
\end{definition}

\begin{definition}[{\sc Directed Steiner Tree Enumeration}]
    Given a directed graph $D = (V, E)$, a root $r \in V$, and a terminal set $W \subseteq V \setminus \{r\}$, the problem asks to enumerate all the minimal subgraphs of $D$ in which there is a directed path from $r$ to $v \in W$.
\end{definition}

\begin{definition}[{\sc Group Steiner Tree Enumeration}]
    \revised{Given an undirected graph $G = (V, E)$ and terminal sets $\mathcal W = \set{W_i \subseteq V: 1 \le i \le s}$, the problem asks to enumerate all minimal group Steiner trees of $(G, \mathcal W)$.}
\end{definition}

\revised{
    As the following problem is slightly different from the other problems, we need further definitions.
    Let $W \subseteq V$ be a terminal set. 
    A subgraph of $G$ is called an \emph{induced Steiner subgraph} of $(G, W)$ if it is a Steiner subgraph of $(G, W)$ and induced by some subset of vertices in $G$.
    An induced Steiner subgraph of $(G, W)$ induced by $U \subseteq V$ is \emph{minimal} if $G[U']$ is not an induced Steiner subgraph of $(G, W)$ for every proper subset $U'$ of $U$. 
}

\begin{definition}[{\sc Induced Steiner Subgraph Enumeration}]
    Given an undirected graph $G = (V, E)$ and a terminal set $W$, the problem asks to enumerate all the minimal induced Steiner subgraphs of $(G, W)$.
\end{definition}

\section{Directed \texorpdfstring{$s$}{s}-\texorpdfstring{$t$}{t} path enumeration revisited}\label{sec:path:enum}
Enumeration of all $s$-$t$ paths \revised{in undirected graphs} is one of the most famous and classical enumeration problems, which is indeed a special case of {\sc Steiner Tree Enumeration}.
There are several efficient enumeration algorithms for this problem~\cite{DBLP:conf/soda/BirmeleFGMPRS13,Read:Tarjan:Networks:1975,Yen:MS:1971,Johnson:SIAM:1975}.
Our proposed algorithms use an $s$-$t$ path enumeration algorithm as a subroutine.
\revised{
    Note that Read and Tarjan~\cite{Read:Tarjan:Networks:1975} gave an amortized $\order{n + m}$ time and $\order{n + m}$ space algorithm for enumerating $s$-$t$ paths in undirected graphs. It is easy to extend this result to directed graphs and turn into an $\order{n + m}$ delay bound.
    However, we cannot use their algorithm as a black box in {\sc Steiner Tree Enumeration} since this may yield the space bound \revised{$\order{\size{W}(n + m)}$, where $\size{W}$ is the number of terminals}.
}
To reduce \revised{$\order{\size{W}(n + m)}$} space to $\order{n + m}$ space, we need to dive into the details of their algorithm and modify it, which is described below.

Let $D = (V, E)$ be a directed graph and 
let $\mathcal P(s, t, D)$ be the set of all directed $s$-$t$ paths in $D$.
The idea of the algorithm is as follows.
We initially compute an arbitrary path $Q = (v_1, \ldots, v_k)$ with $v_1 = s$ and $v_k = t$ in $D$ and output it.
Then, the entire enumeration problem $\mathcal P(s, t, D)$ can be partitioned into subproblems $\mathcal P_i$ for $1 \le i < k$, each of which requires to enumerate all \revised{directed $s$-$t$ paths} that contains $(v_1, \ldots, v_i)$ as a subpath and does not contain edge $(v_i, v_{i+1})$.
Since every path in $\mathcal P(s, t, D)$ is either $Q$ or enumerated in exactly one of these subproblems $\mathcal P_i$, the algorithm correctly enumerates all the paths in $\mathcal P(s, t, D)$.
To solve each subproblem $\mathcal P_i$, it suffices to enumerate all $v_i$-$t$ paths in $D[V \setminus \{v_1, \ldots, v_{i-1}\}] - (v_i, v_{i+1})$.

The algorithm based on this branching strategy is given in \Cref{algo:enum:stpath}.
\revised{
    We assume that for each vertex $v$, the outgoing edges incident to $v$ are totally ordered.
    Given this, for two edges $e, f$ incident to $v$, we denote by $e \prec_v f$ if $e$ is smaller than $f$ in this order and by $e \preceq_v f$ if either $e = f$ or $e \prec_v f$. 
    The algorithm first enumerates directed paths $Q^0, Q^1, \ldots, Q^p$ starting from $s$ such that the edges incident to $s'$ in these paths are distinct. 
    This can be done by \FindSTP with the main loop of \EnumSTP.
    The subroutine \FindSTP{$D', s', t, e, f$} computes a directed $s'$-$t$ path $Q^j$ in $D'$ that avoids $e$ and all outgoing edges $f'$ incident to $s'$ with $f' \preceq_{s'} f$.
    As there can be multiple choices of these directed paths, we choose a path whose edge incident to $s'$ is ``smallest'' among all possible such paths.
    This implies that if $D'$ has an $s'$-$t$ path containing an outgoing edge $(s', v)$, there is also a path $Q^j$ containing it, which is enumerated as above. 
    Then, the algorithm enumerates subproblems for each path $Q^j$ and makes recursive calls. 
}
\revised{On each recursive call,} for a directed $s$-$s'$ path $P$ and a directed $s'$-$t$ path $Q^j$,
\revised{we output $P \circ Q^j$ as a solution, where 
$P \circ Q^j$ is the directed $s$-$t$ path obtained by concatenating $P$ and $Q^j$ with connection $s'$.}

To enumerate all directed $s$-$t$ paths, call \EnumSTP{$(s), \bot, 0, \revised{t}$}.
\revised{
    In what follows, let $\mathcal T$ be the rooted tree generated by the execution of \Cref{algo:enum:stpath} whose root corresponds to \EnumSTP{$((s), \bot, 0, t)$}, that is, this tree structure has a node for each call of \EnumSTP.  
    We call $\mathcal T$ the \emph{enumeration tree}.
}

\begin{lemmarep}
\revised{
    \Cref{algo:enum:stpath} enumerates all directed $s$-$t$ paths without duplication.
    }
\end{lemmarep}
\begin{proof}
\revised{
    Let $\mathcal P(s, t, D)$ be a collection of directed $s$-$t$ paths in $D$. 
    Since every path outputted by \Cref{algo:enum:stpath} is the concatenation of an $s$-$s'$ path and an $s'$-$t$ path, 
    the algorithm only outputs an $s$-$t$ path in $D$.
    Thus, we show that \Cref{algo:enum:stpath} outputs all paths in $\mathcal P(s, t, D)$ without duplication. 
    
    We first prove that the algorithm outputs every path $R\in \mathcal P(s, t, D)$.
    Suppose for contradiction that a path $R = (u_1, \ldots, u_k)$ is not output by the algorithm.
    Let $P = (u_1, \ldots, u_{k'})$ be the maximal subpath of $R$ such that $\EnumSTP(P, e, d, t)$ is called during the execution of the algorithm.
    Such a path $P$ is well-defined since we initially call $\EnumSTP((s), \bot, 0, t)$.
    The edge $e$ does not belong to $R$ due to the maximality of $P$. 
    Moreover, as the main loop in \EnumSTP with \FindSTP enumerates $u_{k'}$-$t$ paths $Q^0, Q^1, \ldots, Q^p$, exactly one of them, say $Q^j$, uses edge $(u_{k'}, u_{k'+1})$ as the outgoing edge incident to $u_{k'}$.
    If $P \circ Q_j = R$, we are done.
    Otherwise, there is an edge $(u_{k''}, u_{k'' + 1})$ in $Q^j$ that does not belong to $R$.
    We indeed call \EnumSTP{$P \circ Q^j_i, (u_{k''}, u_{k''+1}), d+1, t$} for appropriate $i$, which contradicts the maximality.
    
    We show that the output of the algorithm has no duplication. 
    Let $L_1$ and $L_2$ be a pair of leaf nodes in the enumeration tree $\mathcal T$ and 
    $X$ be the lowest common ancestor of $L_1$ and $L_2$.
    The algorithm generates child nodes of $X$ by adding distinct edges to the subpath on $X$.
    Thus, the solutions on $L_1$ and $L_2$ must be distinct and the algorithm has no duplication.
}
\end{proof}
\DontPrintSemicolon
\begin{algorithm}[t]
    \SetAlCapSkip{5ex}
    \SetKwRepeat{Do}{do}{while}%
    \SetAlgoLined
    \AlgInput{$P$: a directed $s$-$s'$ path, \revised{$e$: an edge cannot be used for extending $P$}, $d$: depth of recursion. }
    \revised{
        \Procedure{\EnumSTP{$P, \revised{e}, d, \revised{t}$}}{
            $Q^0 = (v^0_1, \ldots, v^0_k) \gets$ \FindSTP{$D[V \setminus (V(P) \setminus \set{s'})], s', t, e, \bot$}\label{algo:stpath:find:path:1}\;
            $j \gets 0$\;
            \While{$Q^j \neq \bot$ \label{algo:while}}{
                \lIf{$d$ is even\label{algo:stpath:even}}{
                    Output $P \circ Q^j$ \label{algo:stpath:even:output}
                }
                \For{$i = k - 1, \ldots, 2$}{\label{algo:stpath:for}
                    Let $Q^j_i = (v^j_1, \ldots, v^j_i)$ be a subpath of $Q^j$\; \label{alg:stpath:subpath:get}
                    \If{$D[V \setminus (V(P\circ Q^j_i) \setminus \set{v^j_i})] - (v^j_i, v^j_{i + 1})$ has a directed $v^j_i$-$t$ path}{
                        \EnumSTP{$P \circ Q^j_i, \revised{(v^j_i, v^j_{i+1})}, d+1, \revised{t}$} \label{alg:stpath:rec}
                    }
                }
                \lIf{$d$ is odd\label{algo:stpath:odd}}{
                    Output $P \circ Q^j$ \label{algo:stpath:odd:output}
                }
                $Q^{j + 1} = (v^{j+1}_1, \ldots, v^{j+1}_k) \gets$
                \FindSTP{$D[V \setminus (V(P) \setminus \set{s'})], s', t, e, (v^{j}_1, v^{j}_2)$}\label{algo:stpath:find:path:2}\;
                $j \gets j + 1$\;
            }

    }
        \Procedure{\FindSTP{$D', s', t, e, f$}}{
                \lIf{$e \neq \bot$}{Remove $e$ from $D'$}
                \If{$f \neq \bot$}{
                    Remove outgoing edges $f'$ incident to $s'$ with $f' \preceq_{s'} f$ from $D'$\; 
                }
                Compute a directed $s'$-$t$ path $Q = (u_1, \ldots, u_k)$ in $D'$ such that there is no directed $s'$-$t$ path $Q' = (u'_1, \ldots, u'_{k'})$ with $(u'_1, u'_2) \prec_{s'} (u_1, u_2)$ in $D$\;
                \lIf{no such path $Q$ exists}{ \Return $\bot$}
                \Return $Q$\;  
            }
        }
    \caption{
                A linear-delay enumeration algorithm for directed $s$-$t$ path enumeration. 
            } 
    \label{algo:enum:stpath}
\end{algorithm}

\revised{
    We next consider the time complexity of the algorithm. 
    We can easily confirm that \FindSTP runs in $\order{n+m}$ time. 
    For each path $Q$ computed by \FindSTP, we output a solution $P \circ Q$.
    To achieve $\order{n + m}$ delay, we need to enumerate all children in $\order{n + m}$ delay at each node $X$ in the enumeration tree $\mathcal T$.
}
Let $P$ be an $s$-$s'$ path and $Q = (v_1,  \ldots, v_k)$ be an $s'$-$t$ path such that $P \circ Q$ is an $s$-$t$ path in $D$.
For each $\revised{2} \le i \le k$, we let $Q_i = (v_1, \ldots, v_i)$.
If we can enumerate all subpaths $Q_i$ such that $D[V \setminus (V(P \circ Q_i) \setminus \set{v_i})]$ contains a directed $v_i$-$t$ path avoiding $(v_i, v_{i + 1})$ in $\order{n + m}$ delay, then \revised{we can enumerate all children in $\order{n + m}$ delay as well.}
Given this, we say that $Q_i$ is \emph{extendible with $P$} if there is a directed $v_i$-$t$ path in $D[V \setminus (V(P \circ Q_i) \setminus \set{v_i})] - (v_i, v_{i + 1})$.
\revised{The following lemma is crucial for efficiently obtaining such subpaths $Q_i$. }

\begin{lemmarep}\label{lem:stpath:children}
    Let $Q_i$ be a subpath of $Q$ that is extendible with $P$.
    Then, we can, in $O(n + m)$ time, either find the largest index $i^*$ with $i^* < i$ such that $Q_{i^*}$ is extendible with $P$ or determine no such \revised{index} exists.
\end{lemmarep}
\begin{proof}
    Let $2 \le j \le i$.
    To compute the reachability from $v_j$ to $t$ in $D_j = D[V \setminus (V(P \circ Q_j) \setminus \set{v_j})] - \revised{(v_j, v_{j + 1})}$ efficiently, we use a boolean value $r(j, u)$ for each vertex $u$ in $D_j$ defined as follows\footnote{The role of $r(i, u)$ is the same as $d(\cdot)$ in \cite{Read:Tarjan:Networks:1975} but we slightly modify the notion for the expository purpose. }.
    We set $r(j, u)$ to true if and only if $D_j$ has a directed path from $u$ to $t$. 
    For $j = i$, we can obtain all values $r(j, \cdot)$ in $\order{n + m}$ time using a standard graph search algorithm.
    
    We consider how to obtain $r(j, u)$ from $r(j + 1, u)$ for $j < i$. 
    Let $V_j$ and $E_j$ be the set of vertices and the set of edges in $D_j$, respectively.
    Note that $V_{j+1} \subseteq V_{j}$ and $E_{j+1} \subseteq E_{j}$. 
    This implies that $r(j, u)$ is true, then $r(j', u)$ is also true for any $j' < j$. 
    
    We set the value of $r(j, v)$ for any vertex $v$ in $D_j$ as follows. 
    If $v \notin V_{j+1}$, then $r(j+1, v)$ is false. 
    Let $F$ be the set consisting of edges $(v, w)$ in $E_{j} \setminus E_{j+1}$ such that $r(j+1, v)$ is false and $r(j+1, w)$ is true. 
    In the following, we use $F$ as a program variable.
    For each edge $(v, w)$ in $F$, we set $r(j, v)$ to true, remove $(v, w)$ from $F$, and add all incoming edges $(x, v)$ of $v$ to $F$ if $r(j+1, x)$ is false. 
    We repeat this until $F$ is empty.  
    
    We show the correctness of this update procedure. 
    For any vertex $u$ in $D_j$ such that $r(j, u)$ is true, if $r(j+1, u)$ is false, then 
    all the paths from $u$ to $t$ must contain at least one edge in the initial set of $F$. 
    Let $k$ be the shortest distance from $u$ to the closest vertex $w$ in $D_j$ such that $r(j+1, w)$ is true. 
    If $k = 1$, by following edges in the initial set of $F$, we can set the correct value to $r(j, u)$. 
    In addition, if we add an edge to $F$, the tail is reachable to $t$. 
    Thus, for any $k$, we can correctly update the values of $r$ by the transitivity of the reachability to $t$.  
    
    Each edge in $D$ is added to $F$ at most once. 
    Moreover, we maintain the reachability $r(\cdot, \cdot)$ in an array of length $n$ since $r(j, u)$ is true only if $r(j', u)$ is true for all $j' \le j$.
    Thus, this procedure finds $i^*$ in linear time and linear space. 
    %
    %
    %
\end{proof}

We next consider the space complexity of the algorithm.
Let $X$ be a node of $\mathcal T$ and $P_X$ be a directed $s$-$t$ path outputted in $X$. 
To reduce the space complexity of our algorithms for {\sc Steiner Tree Enumeration} and variants,
the important observation is that the total space for storing some information on each node for successive recursive calls is $\order{\size{P_X}}$, 
see the appendix for the details.
The algorithm can be applied to undirected graphs by simply replacing each undirected edge with two directed edges with opposite directions.  
We should remark that Birmel{\'{e}}~\etal proposed an efficient enumeration algorithm for undirected $s$-$t$ paths running in $\order{\sum_{P \in \mathcal P(s, t, G)}\size{P}}$ total time~\cite{DBLP:conf/soda/BirmeleFGMPRS13}. 

\begin{theoremrep}\label{thm:path:linear}
    \Cref{algo:enum:stpath} enumerates all $s$-$t$ paths of a directed (undirected) graph in $\order{n + m}$ delay and $\order{n + m}$ space.  
\end{theoremrep}
\begin{proof}
    We first consider the space complexity.
    We traverse the enumeration tree in a depth-first manner and then need to store some information on each node for successive recursive calls.
    For each node $X$ and its child $Y$ whose associated instances are \revised{$(P, e, d, t)$ and $(P \circ Q^j_i, (v^j_i, v^j_{i+1}), d + 1, t)$}, respectively, 
    we store the subpath $Q^j_i$, the three directed edges $e$, $(v^j_i, v^j_{i+1})$, and $f$, and the loop variables $i, j$.
    Note that the entire path $P \circ Q^j_i$ and the depth $d$ are maintained in a global memory.
    To compute the next sibling $Z$ of $Y$, we need to restore path $P$ and $Q^j$.
    Since the entire path $P \circ Q^j_i$ is stored in a global memory, we can compute $P$ from $Q^j_i$ in $\order{n + m}$ time. 
    From the restored path $P$ and two directed edges $e = (v^j_i, v^j_{i + 1})$ and $f$,
    $Q^j$ can be recomputed in $\order{n + m}$ time as well \revised{since we use the fixed directed path finding algorithm \FindSTP. }
    Paths $Q^j_i$ are edge-disjoint at any nodes having the ancestor-descendant relation. 
    Thus, the total space for storing information in successive recursive calls on a node $X$ is $\order{\size{P_X}}$
    and the space complexity of the algorithm is $\order{n + m}$, where $P_X$ is a directed $s$-$s'$ path on $X$.

    We next discuss the delay of the algorithm.
    We can compute paths $Q^0, Q^1, \ldots, Q^p$ in $\order{n + m}$ delay.
    For each path $Q^j$, by \Cref{lem:stpath:children}, we can compute the next sibling in $\order{n + m}$ time. 
    Moreover, we output a solution in a pre-order (resp. post-order) manner if the depth of the node is even (resp. odd)\footnote{This technique is known as the alternating output method~\cite{Uno::2003}.}.
    \revised{Since each node output at least one solution, }
    we output at least one solution between three consecutive nodes in the depth-first search traversal of $\mathcal T$.
    Thus, the delay of the algorithm is $\order{n + m}$.
\end{proof}

\Cref{algo:enum:stpath} can be extended to the one for enumerating paths between two disjoint subsets $S$ and $T$ of $V$.
Let $D$ be a directed graph. 
We remove all edges directed from $v \in T$ and directed to $v \in S$. 
We then add a vertex $s$ to $D$ and an edge $(s, v)$ for each $v \in S$.
Similarly, we also add a vertex $t$ to $D$ and an edge $(v, t)$ for each $v \in T$.
Then, by enumerating all $s$-$t$ paths and removing $s$ and $t$ from these paths, we obtain all $S$-$T$ paths in the original graph.

\section{Enumerating minimal Steiner trees via branching}\label{sec:mst}
In this section, we give efficient algorithms for {\sc Steiner Tree Enumeration} by incorporating the path enumeration algorithm described in the previous section.
To explain our idea clearly, we first give a simple polynomial-delay and polynomial-space enumeration algorithm.
Then, we give a linear-delay algorithm with a non-trivial analysis.

\subsection{Polynomial-delay enumeration of minimal Steiner trees}
Let $G = (V, E)$ be an undirected graph and let $W \subseteq V$ be terminals.
Our enumeration algorithm, given a graph $G$, $W \subseteq V$, and a subgraph $T$ of $G$,  enumerates minimal Steiner trees of $(G, W)$ that contain $T$ as a subgraph.
In this section, we assume that $G$ is connected.
Moreover, during the execution of our algorithm, $T$ always forms a tree whose leaves are all terminal.
We call such a tree $T$ a \emph{partial Steiner tree of $(G, W)$}.
Note that every leaf of $T$ is terminal but some terminals may not be contained in $T$.
Then, we have the following easy but key lemma for partial Steiner trees. 

\begin{lemma}\label{lem:st:extension}
    If $T$ is a partial Steiner tree of $(G, W)$, then there is a minimal Steiner tree of $(G, W)$ that contains $T$ as a subgraph.
\end{lemma}
\begin{proof}
    Let $T$ be a partial Steiner tree of $(G, W)$.
    As $G$ is connected, there is a spanning tree $T'$ of $(G, W)$ containing $T$ as a subgraph.
    By greedily removing non-terminal leaves from $T'$, we have a Steiner tree of $(G, W)$ containing $T$ as a subgraph whose leaves are all terminal, which is a minimal Steiner tree of $(G, W)$ by \Cref{prop:st:char}.
    Moreover, as every leaf of $T$ is terminal, $T$ is contained in this minimal Steiner tree.
\end{proof}

\DontPrintSemicolon
\begin{algorithm}[t]
    \SetAlCapSkip{0ex}
    \caption{A polynomial-delay algorithm to enumerate all minimal Steiner trees. 
    Note that the root recursive call is \texttt{E-MST($G, W, (\set{w_0}, \emptyset)$)}, where $w_0$ is an arbitrary terminal. }
    \label{algo:enum:st}
    \AlgInput{$G$: a graph, $W$: a set of terminals, $T$: a partial Steiner tree of $(G, W)$}
    \Procedure{\EnumMST{$G, W, T$}}{
        \lIf{$T$ is a minimal Steiner tree of $(G, W)$} {
            Output $T$
        }
        \Else{
        Let $w$ be an arbitrary terminal with $w \notin V(T)$\; \label{algo:enum:st:pick:terminal}
        \For{each $V(T)$-$w$ path $P$ in $G$}{ \label{algo:enum:st:child}
            \EnumMST{$G, W, T + E(P)$}
        }}
    }
\end{algorithm}

Our polynomial-delay enumeration algorithm is shown in \Cref{algo:enum:st}. 
The algorithm is based on branching and defines a rooted tree whose root corresponds to the initial call \EnumMST{$G, W, (\set{w_0}, \emptyset)$}, where $w_0$ is an arbitrary terminal in $W$.
We call this rooted tree an \emph{enumeration tree}.
In each recursion step, fix a terminal $w \in W \setminus V(T)$ and generate new partial Steiner tree $T + E(P)$ for every $V(T)$-$w$ path $P$ in $G$.
By~\Cref{lem:st:extension}, every partial Steiner tree generated by \Cref{algo:enum:st} is a subgraph of some minimal Steiner tree of $(G, W)$, assuming that $G$ is connected.
This implies that the algorithm outputs a minimal Steiner tree of $(G, W)$ at every leaf node in the enumeration tree.
The correctness of the algorithm is given in the following lemma.

\begin{lemma}\label{lem:st:cor}
    Algorithm~\ref{algo:enum:st} enumerates all minimal Steiner trees of $(G, W)$ without duplication. 
\end{lemma}
\begin{proof}
    Let $X$ be a node in the enumeration tree made by the algorithm with initial call \EnumMST{$G, W, (\{w_0\}, \emptyset)$} for some $w_0 \in W$.
    Suppose that the partial Steiner tree associated with $X$ is $T$.
    Let $\mathcal S(X)$ be the set of minimal Steiner trees $T^*$ of $(G, W)$ associated with $X$ that satisfy $E(T) \subseteq E(T^*) \subseteq E(G)$.
    First, we show that the algorithm outputs all minimal Steiner trees in $\mathcal S(X)$ at a leaf of the enumeration tree that is a descendant of $X$.
    Suppose that $X$ is an internal node in the enumeration tree as otherwise, we are done.
    Let $T^*$ be an arbitrary minimal Steiner tree in $\mathcal S(X)$.
    Let $w$ be a terminal with $w \notin V(T)$.
    Since $T^*$ is a Steiner tree of $(G, W)$, there is a unique $V(T)$-$w$ path $P$ in $T^*$.
    Clearly, $T + E(P)$ is a partial Steiner tree of $(G, W)$, there is a child of $X$ whose associated instance is $(G, W, T + E(P))$ as we branch all possible $V(T)$-$w$ paths in \Cref{algo:enum:st}.
    Since $T + E(P)$ is a subtree of $T^*$, by inductively applying this argument, $T^*$ is output at some leaf node in the enumeration tree.
    
    Next, we show that the algorithm does output all minimal Steiner trees of $(G, W)$ without duplication.
    Suppose for contradiction that there is a minimal Steiner tree $T^*$ of $(G, W)$ that is output at two distinct leaf nodes $L_1$ and $L_2$ in the enumeration tree. 
    Let $X$ be the lowest common ancestor node of $L_1$ and $L_2$ and let $T$ be the partial Steiner tree of $(G, W)$ associated to node $X$.
    Let $X_1$ and $X_2$ be the children of $X$ that are ancestor nodes of $L_1$ and $L_2$, respectively.
    As $X_1 \neq X_2$, the partial Steiner trees associated to $X_1$ and $X_2$ contain distinct $V(T)$-$w$ path for some $w \in W \setminus V(T)$.
    This contradicts the uniqueness of the $V(T)$-$w$ path in $T^*$.
\end{proof}

We run the path enumeration algorithm discussed in the previous section at \Cref{algo:enum:st:child} in \Cref{algo:enum:st} and immediately call \EnumMST{$G, W, T + E(P)$} for each path $P$ when it is output by the path enumeration algorithm.
Thus, we have the next theorem for the complexity of \Cref{algo:enum:st}. 

\begin{theorem}\label{theo:st:poly}
    Algorithm~\ref{algo:enum:st} enumerates all minimal Steiner trees of $(G, W)$ in \revised{$\order{\size{W}(n + m)}$} delay and \revised{$\order{\size{W}(n + m)}$} space, 
    where $n$ and $m$ are the number of vertices and edges of $G$, respectively.
\end{theorem}
\begin{proof}
    By using an $\order{n + m}$ delay algorithm for enumerating all $s$-$t$ paths in \Cref{sec:path:enum}, we can enumerate in $\order{n + m}$ delay all children of a node in the enumeration tree.
    Moreover, the depth of the enumeration tree is at most $\size{W}$.
    Hence, the delay of our proposed algorithm is \revised{$\order{\size{W}(n + m)}$}.
    It is sufficient to store the information of recursive calls for backtracking with $\order{n + m}$ space per node.
    Therefore, the algorithm runs in \revised{$\order{\size{W}(n + m)}$} space in total. 
\end{proof}

\subsection{Improving the delay and space bound}
\label{subsec:imp:delay}

\newcommand{\bt}[1]{T_{\texttt{bt}}(#1)}

In this subsection, we show that we can enumerate all the minimal Steiner trees of $(G, W)$ in $\order{n + m}$ delay and $\order{n^2}$ space with $O(nm)$ preprocessing time.
The main obstacle for achieving $\order{n + m}$-delay enumeration is that some internal node in the enumeration tree obtained by \Cref{algo:enum:st} may have \revised{only} one child.
\revised{
    Although each child is generated in $O(n + m)$ time, it is hard to enumerate solutions in linear delay if the number of internal nodes is much larger than the number of leaves. 
    Hence, we need to modify \Cref{algo:enum:st:pick:terminal} in the algorithm so that each internal node in the enumeration tree has at least two children, which implies that the number of internal nodes is at most the number of leaves.
}

To improve the running time bound, we first give an algorithm with amortized $\order{n + m}$ time per solution.
Suppose that every internal node has more than one child node in the enumeration tree defined in the previous section.
The total running time of the algorithm is upper bounded by $\sum_{X} \order{n + m} \cdot \size{ch(X)}$, where the summation is taken over all nodes $X$ in the enumeration tree and $ch(X)$ is the set of children of node $X$.
Since the number of internal nodes is at most the number of leaf nodes and the algorithm outputs exactly one solution at each leaf node, the running time of the algorithm is amortized $\order{n + m}$ time.
Moreover, as we will discuss at the end of this subsection, we can prove that (a modified version of) this algorithm runs in $\order{n + m}$ delay by means of the output queue method due to Uno~\cite{Uno::2003}.

We first ``improve'' the enumeration tree discussed in the previous subsection.
The following lemma is a key to this improvement.

\begin{lemma}\label{lem:Steiner:bridge}
    Let $T$ be a partial Steiner tree of $(G, W)$ and let $w \in W \setminus V(T)$. 
    Let $P$ be a $V(T)$-$w$ path in $G$.
    Then, $P$ is the unique $V(T)$-$w$ path in $G$ if and only if all the edges in $P$ are bridges in $G$.
\end{lemma}

\begin{proof}
    Suppose that there are at least two $V(T)$-$w$ paths, say $P$ and $P'$, in $G$.
    Then, there is a closed walk passing through $P$, $P'$, and some part of $T$, implying that $G$ has a cycle containing at least one edge from $P$, implying that this edge is not a bridge in $G$.
    
    Conversely, suppose there is an edge $e$ in $P$ that is not a bridge of $G$.
    Let $P'$ be a path between the end vertices of $e$ in $G - e$.
    If $V(P') \cap V(T) = \emptyset$, $P - e + E(P')$ is a $V(T)$-$w$ path distinct from $P$, which implies that there are at least two $V(T)$-$w$ paths in $G$.
    Otherwise, we can still find a $V(T)$-$w$ path that is a subpath of $P'$, and hence the lemma follows.
\end{proof}

By \Cref{lem:Steiner:bridge}, using a linear-time bridge enumeration algorithm~\cite{DBLP:journals/siamcomp/Tarjan72}, we can either find a terminal $w \notin V(T)$ such that $G$ has at least two $V(T)$-$w$ paths or conclude that there is no such a terminal in linear time.
To do this, we first compute a minimal Steiner tree $T'$ of $(G, W)$ that contains $T$ as a subgraph.
By~\Cref{prop:st:char}, this can be done in $\order{n + m}$ time by simply finding a spanning tree that contains $T$ and then removing redundant non-terminal leaves.
Then, we can check whether each edge in $E(T') \setminus E(T)$ is a bridge in $G$.
If there is a $V(T)$-$w$ path in $T'$ for some $w \in V(T) \setminus W$ that contains a non-bridge edge in $G$, by~\Cref{lem:Steiner:bridge}, we can conclude that there are at least two $V(T)$-$w$ paths in $G$.
Otherwise, there is no such a terminal, $T'$ is indeed the unique minimal Steiner tree of $(G, W)$ containing $T$ as a subgraph.
In this case, the node associated with $(G, W, T)$ can be considered as a leaf node by modifying the enumeration tree and the unique minimal Steiner tree $T'$ can be computed in $\order{n + m}$ time.
Based on these two cases, we can assume that each internal node of the enumeration tree has at least two children.
We call the enumeration tree defined by this modification the \emph{improved} enumeration tree.

\begin{theorem}\label{theo:st:amortized}
    There is an algorithm that enumerates all minimal Steiner trees of $(G, W)$ in $\order{n + m}$ amortized time per solution with $\order{n + m}$ space.
\end{theorem}
\begin{proof}
    Since we can enumerate all $V(T)$-$w$ paths in $\order{n + m}$ delay and each internal node has at least two children, the total running time of the algorithm is upper bounded by $\order{(n + m)N}$, where $N$ is the number of minimal Steiner trees of $(G, W)$, which yields the claimed running time bound.
    
    We next consider the space complexity. 
    We should note that $G$, $W$, and $T$ are stored as global data structures: We do not make copies on each recursive call.
    $T$ is maintained as edge-disjoint paths in the data structure.
    For each node in the enumeration tree, we store a $V(T)$-$w$ path $P$ of $G$ in the global data structure when calling \EnumMST{$G, W, T + E(P)$} and overwrite $P$ with $P'$ when calling \EnumMST{$G, W, T + E(P')$} for the next $V(T)$-$w$ path $P'$.
    Thus, we can maintain partial Steiner tree $T$ in $\order{n + m}$ space.
    To estimate the entire space consumption, consider a node $X$ associated to instance $(G, W, T)$ in the enumeration tree $\mathcal T$ of {\sc Steiner Tree Enumeration}.
    For each path $P$ output by the $s$-$t$ path enumeration algorithm, 
    \revised{
        we use $\order{\size{P}}$ space, see \Cref{sec:path:enum} for the details.
        Since the total length of paths is equal to the size of a minimal Steiner tree, the algorithm runs in $\order{n + m}$ space.
    }
\end{proof}

To obtain a linear-delay bound, we employ the output queue method by Uno~\cite{Uno::2003}. 
Roughly speaking, since the algorithm traverses the enumeration tree in a depth-first manner, we can see this as an Eulerian tour starting and ending at the root node in the graph obtained by replacing each edge with two parallel edges.
The time elapsed between two adjacent nodes in this tour is upper bounded by $\order{n + m}$.
However, as the algorithm outputs solutions at leaf nodes only, the delay can be \revised{$\Omega(\size{W}(n + m))$}.
The idea of this technique is that we use a buffer of solutions during the traversal.
Since the number of leaf nodes is larger than that of internal nodes, we can ``periodically'' output a solution using this buffer.

\begin{figure}
    \centering
    \includegraphics[width=0.3\textwidth]{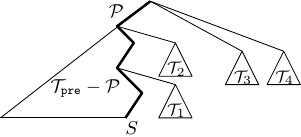}
    \caption{The structure of the improved enumeration tree. The thick line illustrates the path $\mathcal P$ between the root and node $S$ \revised{at which we find the $n$-th solution in the preprocessing phase.}}
    \label{fig:iet}
\end{figure}

We slightly modify the original \revised{method} in~\cite{Uno::2003} to fit our purpose. 
Let $X$ be a node in the improved enumeration tree $\mathcal T$ made by our algorithm.
We say that $X$ is \emph{discovered} when it is visited for the first time and is \emph{examined} when either the parent of $X$ is visited \revised{after visiting $X$ for the last time} or the algorithm halts. 
\revised{We may identify these two cases for each leaf node $X$ since they must occur consecutively.}
Firstly, we find the first $n$ solutions and add them into \revised{a queue $Q$} without outputting these solutions.
Let $\mathcal T_{\tt pre}$ be the subtree of $\mathcal T$ induced by the nodes that are discovered in the preprocessing phase. 
Secondly, after this preprocessing phase, we output solutions conforming to some rules.
Let $\mathcal T_1, \ldots, \mathcal T_\ell$ be the connected components of $\mathcal T - V(\mathcal T_{\tt pre})$, which are indeed subtrees of $\mathcal T$ rooted at some undiscovered nodes (See \Cref{fig:iet}).
We order these subtrees in such a way that for $1 \le i < j \le \ell$, the root of $\mathcal T_i$ is discovered before discovering that of $\mathcal T_j$ during the execution of the algorithm.

\revised{
    To bound the delay of the algorithm, we first discuss the strategy for outputting solutions from $Q$ in $\mathcal T_i$.
    When we traverse $\mathcal T_i$, we output a solution at $X$ from $Q$ according to the following rules: 
    \begin{enumerate}
        \item[\textbf{\textsf{R1.}}] output a solution from $Q$ if $X$ is an internal node, the depth of $X$ is odd in $\mathcal T_i$, and $X$ is examined, 
        \item[\textbf{\textsf{R2.}}] output a solution from $Q$ if $X$ is an internal node, the depth $X$ is even in $\mathcal T_i$, and $X$ is discovered,
        \item[\textbf{\textsf{R3.}}] output the solution found on $X$ if $X$ is a leaf of $\mathcal T_i$ and either $Q$ contains $3n/2$ solutions or $X$ is the third or subsequent leaf child of the parent of $X$ (meaning that there are at least two leaf siblings of $X$ that have been already examined). 
    \end{enumerate}
    For a leaf node $X$, we add a solution that is found on $X$ to $Q$ without outputting it whenever \textbf{R3} does not hold.
}

The entire algorithm is described as follows.
In the preprocessing phase, we compute $n$ solutions by traversing the improved enumeration tree.
Recall that $\mathcal T_{\tt pre}$ is the subtree induced by the nodes discovered in this phase.
Observe that the number of nodes in $\mathcal T_{\tt pre}$ is $O(n)$.
To see this, consider the path $\mathcal P$ between the root and the leaf node $S$ at which we find the $n$-th solution in $\mathcal T$ (see \Cref{fig:iet}).
Clearly, $\mathcal P$ has at most $n$ nodes.
Moreover, as each connected component in $\mathcal T_{\tt pre} - \mathcal P$ is a rooted tree whose internal nodes have at least two children, $\mathcal T_{\tt pre} - \mathcal P$ contains $O(n)$ nodes.
After the preprocessing phase, we proceed with the traversal from $S$ to find the root of $\mathcal T_1$ that is the first discovered node after the preprocessing phase.
During the traversal in $\mathcal T_i$, for $1 \le i \le \ell$, we output solutions as above.
Finally, during the traversal from $S$ to each root of $\mathcal T_i$ in $\mathcal P$, we output a solution from \revised{$Q$} at each odd depth node when it is examined.


\begin{toappendix}
\revised{
    For the sake of simplicity, we assume that $n$ is even. 
    To prove the correctness of the algorithm, we first show that, for each $\mathcal T_i$, the algorithm successfully outputs  solutions from $Q$ according to the rules \textbf{R1}, \textbf{R2}, and \textbf{R3}, assuming that $Q$ contains $r \ge n / 2$ solutions when the root of $\mathcal T_i$ is discovered.
    Moreover, when the root of $\mathcal T_i$ is examined, $Q$ contains at least $r$ solutions, assuming that $n/2 \le r \le n$.
    Fix $\mathcal T_i$.
}
Let $I$ and $L$ be the sets of internal nodes and leaf nodes in $\mathcal T_i$, respectively.
We define a function $\theta: I \to L$ that satisfies the following conditions: 
$\theta$ is injective, $\theta(X)$ is a descendant of $X$ for every $X \in I$, 
\revised{and for $Y \in L$ if $Y$ is the third or subsequent leaf child of a node $X$, then $\theta^{-1}(Y)$ is undefined}.
We say that a function $\theta$ is \emph{consistent} if it satisfies these conditions.
\revised{
    In our proof, we will keep track of solutions in $Q$ based on a consistent function.  
}

\begin{lemma}\label{lem:consistent}
    Let $T$ be a rooted tree such that every internal node has at least two children.
    Then, there is a consistent function from the internal nodes to the leaf nodes in $T$.
\end{lemma}

\begin{proof}
\revised{
    Without loss of generality, we assume that each node in $T$ has at most two leaves as its children.
    We prove the lemma by giving an algorithm to construct a consistent function $\theta$.
    The algorithm traverses $T$ in a depth-first manner.
    When we reach a leaf $\ell$ in $T$, we define $\theta(x) = \ell$, where $x$ is the nearest ancestor of $\ell$ such that $\theta(x)$ has not been defined so far.
    If there is no such node $x$, we do nothing at $\ell$.
    Now, we claim that this function $\theta$ is consistent.
    It is easy to see that for every internal node $x$, $\theta(x)$ is a descendant of $x$ if $\theta(x)$ is defined.
    Thus, we prove that $\theta(x)$ is well-defined for every internal node $x$ in $T$.
    Suppose for contradiction that there is an internal node $x$ such that $\theta(x)$ is not defined in the above algorithm.
    Assume that $x$ is a deepest node satisfying this condition.
    Let $y$ be an arbitrary child of $x$ and let $T_y$ be the subtree of $T$ rooted at $y$.
    Since $\theta(y)$ is defined and the number of internal nodes is strictly smaller than that of leaves in $T_y$,
    there is a leaf node $\ell$ that is not assigned to every internal node in $T_y$.
    By the construction of $\theta$, $\theta(x) = \ell$, contradicting to the assumption.
}
\end{proof}

\revised{
    By the assumption, $Q$ contains at least $r$ solutions for some $n/2 \le r \le n$.
    We call arbitrary $n/2$ of them the \emph{initial solutions} and the remaining $r - n/2$ of them the \emph{excess solutions}.
    In the following, we abuse the queue $Q$ as follows.
    Before starting the transversal of $\mathcal T_i$, we replace each initial solution $S$ in $Q$ with a pair $(S, d)$ for some integer $d$ with $0 \le d \le n-1$ in such a way that $Q$ contains a pair $(S, d)$ for every even $d$ with $0 \le d \le n - 1$.
    The value $d$ is used for proving the correctness of the algorithm.
    This replacement can be done as $Q$ contains at least $n/2$ solutions.
    We also replace each excess solution $S$ with a pair $(S, \ast)$.
    In fact, we do not output these excess solutions during the traversal in $\mathcal T_i$.
    For each node $X$ in $\mathcal T_i$, we denote by $d(X)$ the depth of $X$ in $\mathcal T_i$.
    Then, the rules \textbf{R1}, \textbf{R2}, and \textbf{R3} can be interpreted as the following modified rules:
\begin{enumerate}
    \item[\textbf{\textsf{R'1.}}] output a solution $S$ of the form $(S, d(X))$ in $Q$ if $X$ is an internal node, $d(X)$ is odd, and $X$ is examined, 
    \item[\textbf{\textsf{R'2.}}] output a solution $S$ of the form $(S, d(X))$ in $Q$ if $X$ is an internal node, $d(X)$ is even, and $X$ is discovered, 
    \item[\textbf{\textsf{R'3.}}] output the solution $S$ found on $X$ if $X$ is a leaf of $\mathcal T_i$ and $\theta^{-1}(X)$ is undefined. If $X$ is a leaf, $\theta^{-1}(X)$ is defined, and $Q$ contains $3n/2$ solutions, we do the following.
    If $Q$ contains a solution $S'$ of the form $(S', d(\theta^{-1}(X)))$, then output $S$.
    Otherwise, as $r \le n/2$, there are two solutions $S'$ and $S''$ such that $Q$ contains both $(S', d)$ and $(S'', d)$ for some $d$.
    We replace $(S'', d)$ with $(S, d(\theta^{-1}(X)))$ and output $S''$.
\end{enumerate}
    For leaf node $X$, we add the solution $S$ found on $X$ to $(S, d(\theta^{-1}(X)))$ without outputting it if $Q$ contains less than $3n/2$ solutions.
    Then, the following proposition holds.
    \begin{proposition}\label{prop:output-queue}
        Suppose that $X$ is a leaf in $\mathcal T_i$ and $\theta^{-1}(X)$ is defined, that is, $X$ is either the first or second leaf child of its parent.
        Then, $Q$ contains a pair $(S, d(\theta^{-1}(X)))$, where $S$ is the solution found at $X$, when $X$ is examined.
        Moreover, if $Q$ contains a pair $(S', d)$ for some $0 \le d \le n - 1$ before discovering $X$, then it also contains a pair $(S'', d)$ after examining $X$.
    \end{proposition}
    Observe that these modified rules simulate the original rules: The timing of outputting solutions according to these modified rules are exactly the same with that for the original rules.
    Thus, if we can successfully output solutions according to the modified rules, we do so according to the original rules.
    
    Now, we show that we can output solutions according to the modified rules.
    Let $X$ be an internal node in $\mathcal T_i$.
    Suppose \textbf{R'1} occurs at $X$.
    Then, $Q$ must contain a solution $S$ of the form $(S, d(X))$ since the leaf $\theta(X)$ is already visited before examining $X$ due to the consistency of $\theta$.
    Note that although the solution found at $\theta(X)$ may be different from $S$, by \Cref{prop:output-queue}, $Q$ contains at least one pair $(S, d(X))$.
    Suppose next that \textbf{R'2} occurs at $X$.
    If $X$ is the first depth-$d(X)$ node among all depth-$d(X)$ nodes in $\mathcal T_i$, as $d(X)$ is even, $Q$ contains an initial solution $S$ of the form $(S, d(X))$.
    Otherwise, $X$ is the second or subsequent depth-$d(X)$ node in $\mathcal T_i$.
    Then, $Q$ contains a solution $S$ of the form $(S, d(X))$ since $\theta(Y)$ is already visited at this point, where $Y$ is the depth-$d(X)$ node that is examined immediately before $X$.
    Hence, we can successfully output solutions according to the modified rules.
    
    Recall that we assume $n/2 \le r \le n$. 
    We claim that after traversing $\mathcal T_i$, $Q$ contains at least $r$ solutions.
    
    We observe that after traversing $\mathcal T_i$, $Q$ contains a pair $(S, d)$ for every even $d$ with $0 \le d < n$.
    If $d$ is larger than the height of $\mathcal T_i$, the initial solution $S$ of the form $(S, d)$ that is still in $Q$.
    Otherwise, let $X$ be the depth-$d$ node that is discovered for the last time.
    Then, the solution obtained at $\theta(X)$ is not output in the remaining traversal.
    Thus, $Q$ contains a pair $(S, d)$ for every even $d$ with $0 \le d < n$.
    Moreover, we do not output any excess solutions in $\mathcal T_i$.
    Therefore, $Q$ contains at least $r$ solutions.
    
    
}

\end{toappendix}

\begin{theoremrep}
\label{thm:ste:delay:m}
    \textsc{Steiner Tree Enumeration} can be solved in $\order{n + m}$ delay and $\order{n^2}$ space using $\order{nm}$ preprocessing time. 
\end{theoremrep}

\begin{proof}[Proof of \Cref{thm:ste:delay:m}]
    Since we can generate each child in time $\order{n + m}$, we can move to the next node in the Eulerian tour in time $\order{n + m}$ as well.
    Since $\mathcal T_{\tt pre}$ contains $\order{n}$ node, the preprocessing phase is done in $\order{nm}$ time.
    Let $S$ be the leaf node at which the $n$-th solution is found and let $\mathcal P$ be the path between $S$ and the root of $\mathcal T$.
    From the node $S$ to the root of $\mathcal T_1$, we output a solution at each odd depth node.
    This implies that the delay in this time interval is also $\order{n + m}$, and also for each time interval between two roots of $\mathcal T_i$ and $\mathcal T_{i + 1}$ for $1 \le i < \ell$.
    We output at most $n/2$ solutions at nodes on $\mathcal P$.
    Moreover, if $Q$ contains $r$ ($n/2 \le r \le n$) solutions before traversing $\mathcal T_i$, then it contains at least $r$ solutions after traversing $\mathcal T_i$.
    This implies that we can output solutions using the aforementioned strategy.
    
    To bound the delay in the exploration in $\mathcal T_i$,
    \revised{
        we consider the sequence of internal nodes obtained from the Eulerian tour in $\mathcal T_i$ by removing leaf nodes.
        Let $(X_1, \ldots, X_{10})$ be an arbitrary consecutive subsequence of nodes in the sequence.
        Let us note that as the sequence is obtained from the Eulerian tour, these nodes may not be distinct.
        We show that at least one solution is output at those nodes. 
        If all nodes between $X_{j}$ to $X_{j+3}$ are identical for some $1 \le j \le 7$, then $X_j$ has at least three leaf nodes as its children and we output a solution at one of these children.
        Thus, we assume that the sequence contains at most three consecutive identical nodes.
        If there are only two distinct internal nodes in this sequence, these two nodes appear in an interleaved way, which is impossible in the Eulerian tour obtained from a rooted tree.
        Thus, we can assume that the sequence has at least three internal nodes.
        Let $X_{j_1}$, $X_{j_2}$, and $X_{j_3}$ be three distinct internal nodes such that there is no other internal node between them in the sequence and $1 \le j_1 < j_2 < j_3 \le 10$.
        If all depths of these nodes are distinct, then we output a solution as there is either an odd depth node that is examined (\textbf{R1}) or an even depth node that is discovered (\textbf{R2}). 
        Otherwise, the depth of $X_{j_1}$ and $X_{j_3}$ are identical. 
        Since we examine $X_{j_1}$ and discover $X_{j_3}$, we output a solution.  
        Finally, we consider a minimal consecutive subsequence of nodes in the original Eulerian tour that contains $(X_1, \ldots, X_{10})$ as a subsequence.
        Since leaf nodes do not appear consecutively in this sequence, the length of the sequence is at most $20$ and we can output at least one solution in this sequence.
    }
    Therefore, the delay of this algorithm is $\order{n + m}$.
    
    Finally, we consider the space complexity of the algorithm. 
    \revised{Since $Q$ stores the first $n$ solutions, we use $\order{n^2}$ space for $Q$. }
    The remaining estimation of the space complexity is analogous to \Cref{theo:st:amortized}. 
    Thus, the overall space complexity is $\order{n^2}$.
\end{proof}

We note that this technique can be used when each internal node in the enumeration tree has at least two children.
Thus, we also use this technique to obtain linear delay bounds for other problems, which will be discussed in the next section.

\section{Variants of minimal Steiner trees}
\label{sec:variants}
A similar branching strategy works for other variants of \textsc{Steiner Tree Enumeration}, such as \textsc{Steiner Forest Enumeration}, \textsc{Terminal Steiner Tree Enumeration}, and \textsc{Directed Steiner Tree Enumeration}.
In addition, we can obtain linear-delay enumeration algorithms for these problems.

\DontPrintSemicolon
\begin{algorithm}[t]
    \SetAlCapSkip{0ex}
    \caption{A general description of enumeration algorithms for Steiner problems.}
    \label{algo:enum:general}
    \AlgInput{$G$: a graph, $\mathcal W$: a (family of) terminal set(s), $F$: a partial solution in $G$}
    \Procedure{\EnumSol{$G, \mathcal W, F$}}{
        \lIf{$F$ is a solution}{
            Output $F$
        }
        \Else{
        Let $W$ be a terminal (set) in $\mathcal W$\; \label{algo:enum:general:pick:terminal}
        \For{each valid path $P$ for $(F, W)$}{
            \EnumMST{$G, \mathcal W, F + E(P)$}
        }}
    }
\end{algorithm}

A general form of our algorithm is described as \Cref{algo:enum:general}.
We are given an instance $(G, \mathcal W, F)$, where $\mathcal W$ is a (family of) terminal set(s), and $F$ is a \emph{partial solution}.
Similarly to \Cref{algo:enum:st}, the execution of the algorithm defines a rooted tree whose nodes are associated to triple $(G, \mathcal W, F)$.
We again call this rooted tree an \emph{enumeration tree}.
The algorithm recursively extends $F$ into a larger partial solution $F_P$ by adding a \emph{valid path} $P$ for $F$ and a selected terminal (set) $W$, which can be computed by the $s$-$t$ path enumeration algorithm.
In this section, we focus on \textsc{Steiner Forest Enumeration} as a concrete example of these problems.

\newcommand{\myol}[1]{\makebox[0pt]{$\phantom{#1}\overline{\phantom{#1}}$}#1}
\newcommand{\ReqTer}[2]{\myol{#1}_{#2}}
To specify \Cref{algo:enum:general} for \textsc{Steiner Forest Enumeration}, we define partial solutions and their valid paths.
Recall that for an undirected graph $G = (V, E)$ and a family of terminal sets $\mathcal W = \{W_1, \dots, W_s\}$ with $W_i \subseteq V$, 
a forest $F$ is a \emph{Steiner forest of} $(G, \mathcal W)$ if for any set $W_i$, $F$ contains a path between every pair of vertices in $W_i$. 
In particular, $F$ is a \emph{minimal Steiner forest} of $(G, \mathcal W)$ if there is no proper subgraph of $F$ that is a Steiner forest of $(G, \mathcal W)$. 
Note that when $\size{\mathcal W} = 1$, \textsc{Steiner Forest Enumeration} is equivalent to \textsc{Steiner Tree Enumeration}. 
We assume that each terminal set is contained in a connected component of $G$ as otherwise, there is no minimal Steiner forest of $(G, \mathcal W)$.

In what follows, without loss of generality, we can assume that $\size{W_i} = 2$ for each $W_i$. 
This follows from the observation that if there is a terminal set $W \in \mathcal W$ such that $W = \set{w_1, \ldots, w_k}$ with $k \ge 3$, every Steiner forest of $(G, \mathcal W)$ is a Steiner forest of $(G, \mathcal W')$ with $\mathcal W' = (\mathcal W \setminus \set{W}) \cup \set{\set{w_1, w_2}, \set{w_1, w_3}, \ldots, \set{w_1, w_k}}$ and vice versa. 
Also, from this observation, we can assume that the number of sets in $\mathcal{W}$ is at most $n-1$. 
We note that by this conversion, terminal sets in $\mathcal W$ may intersect.
For each $1 \le i \le s$, we let $W_i = \set{w_i, w'_i}$. 

The next lemma plays an important role for \textsc{Steiner Forest Enumeration}, which enables us to enumerate all the solutions by combining paths, each of which connects a pair of terminals.

\begin{lemma}
\label{lem:msf:is:union:mst}
    $F$ is a minimal Steiner forest of $(G, \mathcal{W})$ if and only if $F$ is a forest consisting of the union of paths $P_1, \dots P_s$ connecting $\set{w_1, w'_1}, \ldots, \set{w_s, w'_s}$, respectively. 
    Moreover, let $\mathcal{P} = (P_1, \dots, P_s)$ and $\mathcal{P}' = (P'_1, \dots, P'_s)$ be two (ordered) sets of $w_i$-$w'_i$ paths $P_i$ and $P'_i$  whose unions are minimal Steiner forests $F$ and $F'$ of $(G, \mathcal W)$, respectively.
    Then, $\mathcal{P} \neq \mathcal{P}'$ if and only if $F \neq F'$.
\end{lemma}
\begin{proof}
    Let $F$ be a forest such that $F$ consists of the union of paths $P_i$ between $w_i$ and $w'_i$ for $1 \le i \le s$.
    Clearly, $F$ is a Steiner forest of $(G, \mathcal W)$.
    Moreover, for any edge $e \in E(P_i)$, $F - e$ has no $w_i$-$w'_i$ path as otherwise $F$ has a cycle. 
    Thus, $F$ is a minimal Steiner forest of $(G, \mathcal W)$.
    
    Conversely, let $F$ be a minimal Steiner forest of $(G, \mathcal W)$. 
    Then, there is a unique path $P_i$ between $w_i$ and $w'_i$ for each $1 \le i \le s$.
    Thus, $F$ contains $\bigcup_{1 \le i \le s}E(P_i)$. 
    Moreover, if $F$ contains an edge $e \in F \setminus \bigcup_{1\le i \le s}E(P_i)$, then this contradicts to the minimality of $F$. 
    Hence, we have $F = \bigcup_{1 \le i \le s}E(P_i)$.     
    
    Clearly, $F \neq F'$ implies $\mathcal P \neq \mathcal P'$ as $F = \bigcup_{1 \le i \le s}E(P_i)$ and $F' = \bigcup_{1 \le i \le s}E(P'_i)$.
    Let $\mathcal P = (P_1, \dots, P_s)$ and $\mathcal P' = (P'_1, \dots, P'_s)$ be as in the statement with $\mathcal P \neq \mathcal P'$.
    Then, there is a pair of $w_i$-$w'_i$ paths $P_i$ and $P'_i$ with $P_i \neq P'_i$.
    Suppose that $F = F'$.
    Since $P_i \neq P'_i$, there are two paths between $w_i$ and $w'_i$ in $F$, which implies that $F$ contains a cycle, a contradiction.
\end{proof}

Now, we define a partial solution and a valid path for \textsc{Steiner Forest Enumeration}.
A forest $F$ of $G$ is called a \emph{partial solution} or, more specifically, a \emph{partial Steiner forest} of $(G, \mathcal W)$ if $F$ is a forest of the form $F = \bigcup_{\set{w_i, w'_i} \in \mathcal W'} E(P_i)$, where $P_i$ is a $w_i$-$w'_i$ path and $\mathcal W'$ is a subset of $\mathcal W$.
Let $F$ be a partial Steiner forest of $(G, \mathcal W)$ and let $W = \set{w, w'} \in \mathcal W$ be a terminal pair such that there is no path between the pair in $F$.
We say that a $w$-$w'$ path $P$ in $G$ is \emph{valid for $(F, W)$} if $F + E(P)$ has no cycles.
Let us note that valid path $P$ may contain edges in $F$.
Clearly, $F + E(P)$ is a partial Steiner forest of $(G, \mathcal W)$ as well.

\begin{lemma}
    \label{lem:msf:extension}
     If $F$ is a partial Steiner forest of $(G, \mathcal W)$, then there is a minimal Steiner forest of $(G, \mathcal W)$ that contains $F$ as a subgraph.
\end{lemma}
\begin{proof}
    Let $F'$ be a maximal forest in $G$ that contains $F$ as a subgraph. 
    By the assumption that every terminal set $W_i$ is contained in a connected component of $G$, $F'$ is a Steiner forest of $(G, \mathcal W)$.
    From $F'$, we repeatedly remove $e \in E(F') \setminus E(F)$ when $F' - e$ is a Steiner forest of $(G, \mathcal W)$.
    Then, we let $F^*$ be the Steiner forest obtained in this way. 
    We show that $F^*$ is a minimal Steiner forest of $(G, \mathcal W)$. 
    
    Suppose that $F^*$ contains an edge $e$ satisfying that $F^* - e$ is a minimal Steiner forest of $(G, \mathcal W)$.
    By the definition of $F^*$, it holds that $e \in F$.
    Since $F$ is a partial Steiner forest, $e$ belongs to a $w$-$w'$ path in $F$ for some $\set{w, w'} \in \mathcal W$.
    If $F - e$ has a $w$-$w'$ path, then $F$ has a cycle, which contradicts to the fact that $F$ is a forest.
\end{proof}

The complete description of \Cref{algo:enum:general} for \textsc{Steiner Forest Enumeration} is as follows. 
We initially call \EnumSol{$G, \mathcal W, (\{w\}, \emptyset)$}, where $w$ is an arbitrary terminal in a terminal set.
Clearly, $(\{w\}, \emptyset)$ is a partial Steiner forest of $(G, \mathcal W)$.
Let $F$ be a partial Steiner forest that is not a Steiner forest of $(G, \mathcal W)$ and let $W = \set{w, w'} \in \mathcal W$ be a terminal pair that is not connected in $F$.
To enumerate all valid paths between $w$ and $w'$ for $(F, W)$, we enumerate all $w$-$w'$ paths $P$ in $G \slash E(F)$.
Since there is a one-to-one correspondence between $E(G) \setminus F$ and $E(G \slash E(F))$, $F + E(P)$ can be seen as a subgraph of $G$.
It is easy to observe that $F + E(P)$ has no cycles, and then it is a partial Steiner forest of $(G, \mathcal W)$, implying that the unique $w$-$w'$ path in $F + E(P)$ is a valid path for $(F, W)$.
The next theorem shows the correctness and running time analysis of the above algorithm.  

\begin{theorem}\label{thm:sf:nm-delay}
    {\sc Steiner Forest Enumeration} can be solved in \revised{$\order{t(n + m)}$} delay and $\order{n + m}$ space\revised{, where $t = \size{\bigcup_{W \in \mathcal W}W}$.}
\end{theorem}
\begin{proof}
    We first show that the algorithm enumerates all the minimal Steiner forests of $(G, \mathcal W)$.
    Let $F^*$ be a minimal Steiner forest of $(G, \mathcal W)$ and $F$ be a partial Steiner forest that is strictly contained in $F^*$.
    By the minimality of $F^*$, there is a terminal pair $W = \set{w, w'}$ that is not connected in $F$.
    Let $P$ be the unique path between $w$ and $w'$ in $F^*$.
    Then, $F + E(P)$ has no cycles, which implies that $P$ is a valid path for $(F, W)$.
    By inductively applying the same argument to $F + E(P)$, we can eventually compute $F^*$.

    Next, we show that the algorithm outputs minimal Steiner forests without duplication.
    Suppose for contradiction that there is a minimal Steiner forest $F^*$ of $(G, \mathcal W)$ that is output at two leaf nodes $L_1$ and $L_2$ in the enumeration tree.
    Let $X$ be the lowest common ancestor of $L_1$ and $L_2$ and let $F$ be the partial Steiner forest associated with $X$.
    Similarly to \Cref{lem:st:cor}, the two children of $X$ that are ancestors of $L_1$ and $L_2$ are associated to distinct partial Steiner forests $F + E(P_1)$ and $F + E(P_2)$ for some distinct valid paths $P_1$ and $P_2$ for $(F, W)$ with some terminal set $W$.
    However, by~\Cref{lem:msf:is:union:mst}, any minimal Steiner forests that respectively contain $F + E(P_1)$ and $F + E(P_2)$ must be distinct, contradicting to the assumption. 
    
    We finally analyze the delay and the space complexity.
    Note that the height of the enumeration tree obtained by our algorithm is at most $n$. 
    Hence, since we use an $\order{n + m}$ delay $s$-$t$ path enumeration algorithm, the delay of our algorithm is \revised{$\order{t(n + m)}$}. 
    At each internal node, we store exactly one valid path for a partial Steiner forest $F$ and, by an analogous argument in \Cref{theo:st:amortized}, the space complexity is $\order{n + m}$.
\end{proof}

In the remaining of this section, we improve the delay complexity of the algorithm in the above theorem with a polynomial-time preprocessing phase as in \Cref{thm:ste:delay:m}. 
To this end, we need to ensure that each internal node of the enumeration tree has at least two children.
Let $X$ be an internal node in the enumeration tree and let $(G, \mathcal W, F)$ be the instance associated with $X$.
Let $W = \set{w, w'} \in \mathcal W$ be a terminal set such that there is no path between $w$ and $w'$ in $F$.
Then, there are at least two valid paths for $(F, W)$ if and only if $X$ has at least two children in the enumeration tree.
By the one-to-one correspondence between $E \setminus E(F)$ and $E(G \slash E(F))$, from every $w$-$w'$ path $P$ in $G \slash E(F)$, we can obtain a unique valid path for $(F, \mathcal W)$.
We note that the graph $G \slash E(F)$ may contain multiedges between two vertices, and then they are not considered as bridges even if removing these edges increases the number of connected components of $G$. 
This correspondence gives the following lemma.

\begin{lemma}\label{lem:sfe:uniqueness}
    Let $F$ be a partial Steiner forest of $(G, \mathcal W)$ that has no $w$-$w'$ path for some terminal pair $W = \set{w, w'} \in \mathcal W$. 
    Let $P$ be a valid path for $(F, W)$.
    Then, $P$ is the unique valid path for $(F, W)$ in $G$ if and only if every edge in $E(P) \setminus E(F)$ is a bridge in $G \slash E(F)$.
\end{lemma}
\begin{proof}
    We first show the if direction. 
    To prove the uniqueness, suppose that there is another valid path $P' \neq P$ for $(F, W)$.
    As $F$ is a forest, we have $E(P) \setminus E(F) \neq E(P') \setminus E(F)$.
    This implies that there are two distinct paths between $w$ and $w'$ in $G \slash E(F)$, contradicting to the fact that every edge on $P^*$ is a bridge in $G \slash E(F)$.

    We next show the other direction. 
    Suppose that $P$ is the unique valid path for $(F, W)$ in $G$.
    Since $P$ is valid, $F + E(P)$ is a forest, and then $P' = P \slash E(F)$ is a $w$-$w'$ path in $G \slash E(F)$. 
    For contradiction, suppose that $P'$ contains a non-bridge edge $e$ in $G \slash E(F)$. 
    Then, there is another $w$-$w'$ path $P''$ in $G \slash E(F)$ that does not contain $e$. 
    Since $F + E(P'')$ is also a partial Steiner forest, the unique path $F + E(P'')$ is a valid path for $(F, W)$, contradicting to the uniqueness of $P$.
    Hence, the statement holds.
\end{proof}

By \Cref{lem:sfe:uniqueness}, 
we can determine whether there exists a pair $W \in \mathcal W$ such that there are at least two valid paths for $(F, W)$ in linear time as follows. 
We could not directly apply a similar idea used in {\sc Steiner Tree Enumeration} to {\sc Steiner Forest Enumeration} as it is not obvious to compute a minimal Steiner forest of $(G, \mathcal W)$ that contains $F$ as a subgraph in linear time.
First, we compute $G' = G \slash E(F)$ in linear time. 
Then, we find the set $B$ of bridges in $G'$ by a linear-time bridge enumeration algorithm~\cite{DBLP:journals/siamcomp/Tarjan72}, and obtain $G'' = G' \slash B$. 
By \Cref{lem:sfe:uniqueness}, there is a unique valid path for $(F, W)$ in $G$ if and only if two terminals in $W$ are identical in $G''$.
Thus, such a pair can be found in $\order{n + m}$ time if it exists. 

Next, we consider the other case, that is, for every terminal set $W \in \mathcal W$ there is exactly one valid path $P$ for $(F, W)$.
In this case, there is a unique minimal Steiner forest $F'$ of $(G, \mathcal W)$ that contains $F$ as a subgraph.
By \Cref{lem:sfe:uniqueness}, $E(P) \setminus E(F)$ is composed of bridges in $G' = G \slash E(F)$.
Let $B$ be the bridges in $G'$. 
From the assumption, any terminal pair $\set{w, w'} \in \mathcal W$ is connected in $F + B$. 
Moreover, $F + B$ forms a forest as $B$ consists of bridges in $G'$. 
However, $F + B$ may not be a minimal Steiner forest of $(G, W)$. 
Thus, to obtain the unique minimal Steiner forest $F'$ from $F + B$, we need to remove redundant bridges in $B$. 
Since we can independently remove redundant bridges from each connected component in $F + B$, we assume that $F+B$ forms a tree and all leaves are terminal.

The naive approach to obtain the unique $F'$ is that we enumerate paths between each terminal pair in $\mathcal{W}$ and take the union of them. 
This approach runs in \revised{$\order{tn}$} time\revised{, where $t = \size{\bigcup_{W \in \mathcal W} W}$.}
Thus, we employ another approach in which we exploit the lowest common ancestor (LCA, for short) of each terminal pair. 
Let $T = F + B$. 
We assume that $T$ is a rooted tree by choosing an arbitrary vertex as its root.
For any pair of vertices $u$ and $v$ in $T$, we denote by $lca(u, v)$ the LCA of $u$ and $v$ in $T$.
The $u$-$v$ path in $T$ can be decomposed into $P(u, lca(u, v))$ and $P(lca(u, v), v)$, where $P(x, y)$ be the unique $x$-$y$ path in $T$.
To compute the LCA for pairs of terminals efficiently, we use a data structure due to Harel and Tarjan~\cite{doi:10.1137/0213024}.
This data structure can be constructed in $\order{n}$ time using $\order{n}$ space and allows us to compute $lca(u, v)$ in $\order{1}$ time for given $u$ and $v$.
Thus, for each terminal pair $\set{w, w'}$ in $T$, we can find the $w$-$w'$ path in $\order{\size{E(P(w, w'))}}$ time.
However, the running time is still quadratic to find all paths between terminal pairs.

To compute the set of edges in $T$ that belong to at least one path between a pair of terminals, we make pairs $(lca(w, w'), w)$ and $(lca(w, w'), w')$ for every terminal pair $\{w, w'\}$ and sort them in the descending order of the height of $lca(w, w')$.
Since there are at most $2n$ pairs and $lca(w, w')$ is an integer between $1$ and $n$, we can sort them in $\order{n}$ time.
Then, for each pair $(a, w)$ processed in this order, we mark all the edges on the path starting from $w$ to its ancestor $a$ and stop marking when a marked edge is found.
Since we do this in the sorted order, all the edges between $w$ and $a$ are already marked.
Thus, this marking process is done in total $\order{n}$ time for all terminal pairs, and, by removing all unmarked edges, we can obtain the unique minimal Steiner forest $F'$ of $(G, \mathcal W)$ that contains $F$ as a subgraph.

\begin{theorem}\label{thm:sfe:delay}
\textsc{Steiner Forest Enumeration} can be solved in $\order{n + m}$ amortized time per solution and $\order{n + m}$ space. 
If we use $\order{nm}$ preprocessing time, this problem can be solved in $\order{m}$ delay with $\order{n^2}$ space.
\end{theorem}
\begin{proof}
    The correctness of the algorithm follows from \Cref{thm:sf:nm-delay}.
    
    Let $X$ be a node in the enumeration tree.
    Let $F$ be a partial Steiner forest of $(G, \mathcal W)$ that is associated with $X$. 
    We can find a terminal pair $W \in \mathcal W$ such that there are at least two valid paths for $(F, W)$ in $\order{n + m}$ time if it exists. 
    In this case, there are at least two children of $X$, each of which can be generated in $\order{n + m}$ delay with the $s$-$t$ path enumeration algorithm.
    Otherwise, we can obtain the unique minimal Steiner forest $F'$ of $(G, \mathcal W)$ in $F + B$ in $\order{n + m}$ time. 
    Since each internal node of this ``improved'' enumeration tree has at least two children,
    the amortized running time of the algorithm is $\order{n + m}$ using $\order{n + m}$ space.
    By applying the output queue technique as in \Cref{theo:st:poly}, we have the $\order{n + m}$ delay bound with $\order{n^2}$ space.
\end{proof}


\begin{toappendix}
\subsection{Terminal Steiner trees}
In this subsection, we give a linear-delay and linear-space enumeration algorithm for {\sc Terminal Steiner Tree Enumeration}.
The essential difference from {\sc Steiner Tree Enumeration} is that every solution does not contain any terminals as an internal vertex, which can be easily handled.
The analysis is almost the same as the one for {\sc Steiner Tree Enumeration}.
When $\size{W}$ is equal to two, the problem is identical to the $s$-$t$ path enumeration problem.
In this case, we can enumerate all minimal terminal Steiner tree in $\order{n + m}$ delay and $\order{n + m}$ space as proved in \Cref{thm:path:linear}.
Thus, in what follows, assume that $\size{W} > 2$.

Let $G = (V, E)$ be an undirected graph and let $W \subseteq V$ be a set of terminals.
Recall that a Steiner tree $T$ of $(G, W)$ is called a terminal Steiner tree if every terminal in $W$ is a leaf in $T$.
A terminal Steiner tree $T$ is called a \emph{minimal terminal Steiner tree} of $(G, W)$ if every proper subgraph of $T$ is not a terminal Steiner tree of $(G, W)$.
It is straightforward to verify the following proposition.
\begin{proposition}\label{prop:tst:char}
    $T$ is a minimal terminal Steiner tree of $(G, W)$ if and only if it is a terminal Steiner tree whose leaves are all terminal.
\end{proposition}
We first characterize the condition that there is at least one terminal Steiner tree of $(G, W)$.

\begin{lemmarep}\label{lem:tst:corner}
    Suppose that $W$ has at least three terminals.
    Then, there is a terminal Steiner tree of $(G, W)$ if and only if there is a component $C$ in $G[V \setminus W]$ with $W \subseteq N_G(C)$.
    Moreover, every terminal Steiner tree of $(G, W)$ has no edges between terminals and edges in a component $C$ in $G[V \setminus W]$ with $W \setminus N_G(C) \neq \emptyset$.
\end{lemmarep}
\begin{proof}
    Clearly, if there is a terminal Steiner tree $T$ of $(G, W)$, then $G[V \setminus W]$ has a component $C_T$ such that $C_T$ contains $T$ and $W \subseteq N_G(C_T)$ holds.
    For the converse, suppose that there is a component $C$ in the statement.
    Then, we take a spanning tree $T$ in $C$ and add a leaf edge between $w$ and a vertex in $V(T)$ for each $w \in W$.
    This can be done by the fact that $W \subseteq N_G(C)$, and hence the obtained graph is a terminal Steiner tree.
    
    Suppose that there is a terminal Steiner tree $T$ that has an edge $e$ between terminals or in a component $C$ of $G[V \setminus W]$ with $W \setminus N_G(C) \neq \emptyset$.
    Since there is a terminal $w$ that is neither an end vertex of $e$ nor contained in $N_G(C)$, the path between one of the end vertex of $e$ and $w$ must pass through some terminal $w' \neq w$, contradicting to the fact that $T$ is a terminal Steiner tree. 
\end{proof}

By~\Cref{lem:tst:corner}, we assume that $W$ is an independent set of $G$ and every component $C$ in $G[V \setminus W]$ satisfies $W \subseteq N_G(C)$.

Now, we define partial solutions and valid paths for {\sc Terminal Steiner Tree Enumeration}. 
We say that a tree $T$ in $G$ is a \emph{partial solution}, or more specifically, a \emph{partial terminal Steiner tree} of $(G, W)$ if either (1) $T$ is the empty graph or (2) every leaf of $T$ is terminal and there is a connected component $C_T$ of $G[V \setminus W]$ such that $W \subseteq N(C_T)$ and $T \subseteq E(G[C_T \cup W])$.
Suppose that $T$ is a partial terminal Steiner tree that is not a terminal Steiner tree of $(G, W)$.
Let $w$ be a terminal in $W \setminus V(T)$. 
Then, a \emph{valid path} $P$ for $(T, w)$ is defined as: (1) a $w$-$w'$ path in $G$ for fixed $w' \in W \setminus \set{w}$ if $T$ is the empty graph, or (2) a $(V(T) \setminus W)$-$w$ path in $G[C_T \cup W]$.
By the assumption that $W$ is an independent set and every component $C$ in $G[V \setminus W]$ satisfies $W \subseteq N_G(C)$, $T + E(P)$ is a partial terminal Steiner tree of $(G, W)$. 
The following lemma is essential to show that {\sc Terminal Steiner Tree Enumeration} can be solved by our strategy.

\begin{lemmarep}\label{lem:tst:all}
    If $T$ is a partial terminal Steiner tree of $(G, W)$, then there is a minimal terminal Steiner tree of $(G, W)$ that contains $T$ as a subgraph. 
\end{lemmarep}
\begin{proof}
    From the definition of a partial terminal Steiner tree, either $T$ is the empty graph or there is a component $C_T$ of $G[V \setminus W]$ such that $W \subseteq N_G(C)$ and $E(T) \subseteq E(G[C_T \cup W])$.
    Since $T$ has no cycles, there is a spanning tree of $C_T$ that contains all the edges in $E(T) \cap E(C_T)$. 
    As $W \subseteq N_G(C_T)$, this spanning tree can be extended to a terminal Steiner tree of $(G, W)$ by adding an edge $\{v, w\}$ for each $w \in W \setminus V(T)$ with some leaf $v \in N(w) \cap C_T$ and an edge $\{v, w\} \in T$ for $w \in W \cap V(T)$.
    This terminal Steiner tree may have a non-terminal leaf.
    By repeatedly removing such a non-terminal leaf and, by \Cref{prop:tst:char}, we have a minimal terminal Steiner tree $T^*$ of $(G, W)$.
    Since every leaf of $T$ is terminal, every edge in $T$ is not removed in this process.
    Hence, $T^*$ is a minimal terminal Steiner tree containing $T$. 
\end{proof}

Now, we briefly describe an enumeration algorithm for terminal Steiner trees according to \Cref{algo:enum:general}. 
The main idea is similar to the algorithm for {\sc Steiner Tree Enumeration}.
We initially call \EnumSol{$G, W, T_\emptyset$}, where $T_{\emptyset}$ is the empty graph.
Let $T$ be a partial terminal Steiner tree of $(G, W)$.
If $T$ contains all the terminals in $W$, then $T$ is a minimal terminal Steiner tree of $(G, W)$ and we output it.
Otherwise, there is a terminal $w$ not contained in $T$.
By the assumption that $W$ is an independent set of $G$ and every component $C$ in $G[V \setminus W]$ satisfies $W \subseteq N_G(C)$, there is at least one valid path for $(T, w)$.
We extend $T$ by adding a valid path $P$ for $(T, w)$.
To do this, we need to enumerate all valid paths for $(T, w)$.
For case (1) where $T$ is the empty graph, we just enumerate all $w$-$w'$ paths in $G$.
For case (2), we first compute the component $C_T$ in $G[V \setminus W]$ and select a terminal $w$ from $W \setminus V(T)$. 
Then, we enumerate all the valid paths for $(T, w)$ by enumerating all $(V(T) \setminus W)$-$w$ paths in $G[C_T \cup W]$. 
The correctness of the algorithm follows from an analogous argument in \Cref{lem:st:cor}, and its complexity follows from an analysis almost identical to the one in \Cref{thm:sf:nm-delay}.
Thus, we have the following.

\begin{theorem}
    \textsc{Terminal Steiner Tree Enumeration} can be solved in $\order{nm}$ delay and $\order{n + m}$ space. 
\end{theorem}

The remaining of this subsection is devoted to showing a linear-delay algorithm for {\sc Terminal Steiner Tree Enumeration}.
To this end, we improve the enumeration tree so that each internal node has at least two children.

\begin{lemmarep}
\label{lem:tst:uniquness}
    Let $T$ be a non-empty partial Steiner tree of $(G, W)$, let $w$ be a terminal not contained in $T$, and let $P$ be a valid path for $(T, w)$.
    Then, $P$ is the unique valid path for $(T, w)$ in $G$ if and only if every edge in $P$ is a bridge in $G[C_t \cup W]$. 
\end{lemmarep}
\begin{proof}
    Suppose that there are at least two valid paths $P$ and $P'$ for $(T, w)$ in $G$.
    By an analogous argument in \Cref{lem:Steiner:bridge}, there is a cycle in $G[C_t \cup W]$ that contains at least one edge from both paths $P$ and $P'$.
    This implies that every $(V(T) \setminus W)$-$w$ path in $G[C_t \cup W]$ has a non-bridge edge in $G[C_T \cup W]$.
    
    Conversely, if $P$ has a non-bridge edge $e$ in $G[C_T \cup W]$, rerouting $P$ along with a cycle passing through $e$ yields a valid path distinct from $P$ for $(T, w)$.
\end{proof}

From this lemma, we can find either a terminal $w$ such that $G[C_T \cup W]$ has at least two $V(T)$-$w$ paths or a unique minimal terminal Steiner tree of $(G, W)$ that contains $T$ as a subgraph in linear time when $T$ is a non-empty partial Steiner tree of $(G, W)$.
To do this, a similar idea used in {\sc Steiner Tree Enumeration} works well.
First, we enumerate all the bridges in $G[C_T \cup W]$ in linear time using a linear-time bridge enumeration algorithm~\cite{DBLP:journals/siamcomp/Tarjan72}.
Then, we compute an arbitrary minimal terminal Steiner tree $T'$ of $(G, W)$ that contains $T$ as a subgraph in linear time.
This can be done by taking an arbitrary spanning tree that contains $T$ in $G[C_t \cup W]$ and removing non-terminal leaves.
By~\Cref{prop:tst:char}, $T'$ is a minimal terminal Steiner tree of $(G, W)$.
Using the bridges in $G[C_T \cup W]$ and $T'$, by~\Cref{lem:tst:uniquness}, we can ``improve'' the enumeration tree in the sense that each internal node has at least two children except for the root node.
When $T$ is the empty graph, we cannot apply \Cref{lem:tst:uniquness}.
However, this exceptional case can be seen as a ``linear-time preprocessing''.
Hence, by a similar argument to \Cref{thm:sfe:delay},  we can obtain the following theorem. 

\begin{theorem}\label{thm:tst}
    \textsc{Terminal Steiner Tree Enumeration} can be solved in $\order{n + m}$ amortized time per solution and $\order{n + m}$ space. 
    If we allow $\order{n^2}$ space and $\order{nm}$ preprocessing time, this problem can be solved in $\order{n + m}$ delay.
\end{theorem}

\subsection{Directed Steiner trees}
In this subsection, we develop a linear-delay enumeration algorithm for minimal directed Steiner trees. 
Let $D = (V, E)$ be a directed graph with terminal set $W \subseteq V$ and let $r \in V \setminus W$.
Recall that a subgraph $T$ of $D$ is a directed Steiner tree of $(D, W, r)$ if $T$ is a directed tree rooted at $r$ that contains an $r$-$w$ path for each $w \in W$ and $T$ is a minimal directed Steiner tree of $(D, W, r)$ if no proper subgraph of $T$ is a directed Steiner tree of $(D, W, r)$.
\begin{proposition}\label{prop:dst:char}
    $T$ is a minimal directed Steiner tree of $(D, W, r)$ if and only if it is a directed Steiner tree whose leaves are all terminal.
\end{proposition}
Without loss of generality, for any vertex $v \in V$, $D$ has a $r$-$v$ directed path. 
By a similar argument as in the other variants, we say a directed tree $T$ rooted at $r$ is a \emph{partial solution}, or more specifically, a \emph{partial directed Steiner tree} of $(D, W, r)$ if all the leaves in $T$ are terminals.
Let $w$ be an arbitrary terminal not in a partial directed Steiner tree $T$. 
We say that a directed path $P$ is \emph{valid} for $(T, w)$ if $P$ is a directed $V(T)$-$w$ path. 
Then, we observe the following key lemma.

\begin{lemmarep}\label{lem:dst:all}
    If $T$ is a partial directed Steiner tree of $(D, W, r)$, then there is a minimal directed Steiner tree of $(D, W, r)$ that contains $T$ as a subgraph. 
\end{lemmarep}
\begin{proof}
    Let $T$ be a partial directed Steiner tree of $(D, W, r)$. 
    Then, from the assumption that each vertex is reachable from $r$, we can obtain a spanning tree $T'$ containing $T$ by adding paths to vertices not in $T$ 
    so that each vertex in $T'$ can be reachable from $r$. 
    Clearly, $T'$ is a directed Steiner tree of $(D, W, r)$ with root $r$. 
    Repeatedly removing non-terminal leaves from $T'$ yields a minimal directed Steiner tree of $(D, W, r)$ by~\Cref{prop:dst:char}. 
\end{proof}

We now describe our enumeration algorithm for minimal directed Steiner trees. 
We initially call \EnumSol{$D, W, T_0 = (\set{r}, \emptyset)$}.
Let $T$ be a partial directed Steiner tree of $(D, W, r)$. 
On recursive call \EnumSol{$D, W, T$}, if $T$ contains all terminals, we output it.
Otherwise, we find a terminal $w \in W \setminus V(T)$. 
Then, we compute all the valid paths for $(T, w)$ by enumerating $V(T)$-$w$ paths in $D$. 
By the assumption that $w$ is reachable from $r$, there is at least one valid path for $(T, w)$.
For each valid path $P$ for $(T, w)$, we call \EnumSol{$D, W, T + E(P)$}.  
The next theorem directly follows from an analogous argument in \Cref{thm:sf:nm-delay}.

\begin{theorem}
    We can enumerate all minimal directed Steiner trees in $\order{nm}$ delay and $\order{n + m}$ space.
\end{theorem}

In the remaining of this subsection, we give a linear-delay enumeration algorithm for \textsc{Directed Steiner Tree Enumeration}. 
To this end, we ``improve'' the enumeration tree so that each internal node has at least two children.
Once we can make this improvement in linear time for each node in the enumeration tree, we can prove, as in the previous subsections, that the entire algorithm runs in $\order{n + m}$ amortized time and $\order{n + m}$ space. 
To find either a terminal $w \in W \setminus V(T)$ such that there are at least two valid paths for $(T, w)$ or the unique minimal directed Steiner tree of $(D, W, r)$ that contains $T$ as a subgraph, we consider the multigraph $D' = D \slash E(T)$, which is obtained from $D$ by contracting all edges in $T$.
If $D'$ has at least two directed $r_T$-$w$ paths for some terminal $w \in W \setminus V(T)$, then we can immediately conclude that there are at least two valid paths for $(T, w)$.
To see this, let us consider an arbitrary depth-first search (DFS) tree $T'$ in $D'$ starting at node $r_T$ corresponding to the contracted part $T$ in $D'$ and a total order $\prec$ on $V(T')$ determined by a post-order transversal in $T'$.
Let $W' = W \setminus V(T)$ and let $T^*$ be the minimal directed Steiner tree of $(D', W', r_T)$ that is a subtree of $T'$.

\begin{lemmarep}\label{lem:dst:desired:path}
    $D'$ has a minimal directed Steiner tree of $(D', W', r_T)$ distinct from $T^*$ if and only if there exists a pair $u$ and $v$ of distinct vertices with $u \prec v$ in $T^*$ such that $D' - E(T^*)$ has a directed $v$-$u$ path.
\end{lemmarep}
\begin{proof}
    For a directed path $P$, we denote by $t(P)$ the unique sink of $P$.
    
    Suppose first that $D - E(T^*)$ has a directed $v$-$u$ path $P$ such that $u \prec v$.
    As $\prec$ is determined by a post-order transversal in $T'$, either $v$ is an ancestor of $u$ in $T'$, or they have no the ancestor-descendant relationship.
    If $v$ is an ancestor of $u$ in $T'$, then $T^* - E(P') + E(P)$ is a minimal directed Steiner tree of $(D', W', r_T)$ with distinct from $T^*$, where $P'$ is defined to be the maximal subpath in $T^*$ with $t(P') = u$ that has neither $v$ nor any terminals as an internal vertex.
    Otherwise, let $w$ be the lowest common ancestor of $u$ and $v$ in $T^*$.
    Then, $T^* - E(P') + E(P)$ is a minimal directed Steiner tree of $(D', W', r_T)$ with distinct from $T^*$, where $P'$ is defined to be the maximal subpath in $T^*$ with $t(P') = u$ that has neither $w$ nor any terminals as an internal vertex.
    
    Conversely, let $T^{**}$ be a minimal directed Steiner tree of $(D', W', r_t)$ with $T^{**} \neq T^*$.
    Then, there exist two distinct $r$-$w$ paths $P^*$ and $P^{**}$ in $T^*$ and $T^{**}$ for some $w \in W'$, respectively.
    Let $Q$ be the maximal subpath of $P^*$ with $V(Q) \subseteq V(P^*) \cap V(P^{**})$ and $u$ be the source vertex of $Q$.
    Since $w \in V(P^*) \cap V(P^{**})$, $u$ is well-defined.
    Let $Q'$ be the maximal subpath of $P^{**}$ with $t(Q') = u$ such that every internal vertex does not belong to $T^*$ and let $v$ be the source vertex of $Q'$.
    Since $r \in V(T^*) \cap V(P^{**})$, $v$ is well-defined.
    As $P^* \neq P^{**}$, we have $u \neq v$.
    
    Now we claim that $u \prec v$ in $T^*$ and $D' - E(T^*)$ has a directed $v$-$u$ path.
    If $v \prec u$, then either $u$ is an ancestor of $v$ in $T^*$ or there is no directed path from $v$ to $u$ in $D'$, which contradicts to the choice of $u$ and $v$.
    Moreover, $Q'$ is indeed a directed $v$-$u$ path in $D' - E(T^*)$, completing the proof of the lemma.
\end{proof}

By~\Cref{lem:dst:desired:path}, to find a terminal $w$ that has at least two valid paths for $(T, w)$, 
it is sufficient to check whether there is a directed path from a larger vertex to a smaller vertex in $D - E(T^*)$ with respect to $\prec$.
This can be done as follows.
From the largest vertex $v$ with respect to $\prec$, we compute the reachability of each vertex in $D'$ by a standard graph search algorithm.
If there is a vertex $u$ that is reachable from $v$, we are done.
Otherwise, we remove all the vertices reachable from $v$ and repeat the same procedure until we find such a path or the graph is empty.
Since $u$ is reachable from some $v'$ with $u \prec v'$ in the original graph $D'$ if and only if either $u$ is reachable from $v$ in $D'$ or $u$ is reachable from $v'$ in the removed graph, this algorithm works correctly.
Clearly, the algorithm runs in $\order{n + m}$ time.
Hence, we have the following conclusion.

\begin{theoremrep}\label{thm:dst}
    \textsc{Directed Steiner Tree Enumeration} can be solved in amortized $\order{n + m}$ time and $\order{n + m}$ space. 
    If we allow $\order{nm}$ preprocessing time, this problem can be solved in $\order{n + m}$ delay with $\order{n^2}$ space.
\end{theoremrep}
\end{toappendix}

\section{Hardness of internal Steiner trees and group Steiner trees}\label{sec:hardness}
In this section, we discuss some \revised{of the hard variants} of {\sc Steiner Tree Enumeration}.
Recall that a Steiner tree of $(G, W)$ is called an \emph{internal Steiner tree} if every vertex in $W$ is an internal vertex of the tree.
Let $s, t \in V$ be distinct vertices and $W = V \setminus \set{s, t}$.
Then, there is an internal Steiner tree of $(G, W)$ if and only if $G$ has an $s$-$t$ Hamiltonian path, which implies the following theorem.

\begin{theorem}
    Unless P $=$ NP, there is no incremental-polynomial time algorithm for {\sc Internal Steiner Tree Enumeration}.
\end{theorem}

We also remark that, as opposed to {\sc Steiner Tree Enumeration} and several variants discussed above, we show that {\sc Group Steiner Tree Enumeration} is at least as hard as {\sc Minimal Transversal Enumeration}.
Given a hypergraph $\mathcal H = (U, \mathcal E)$, {\sc Minimal Transversal Enumeration} is the problem of enumerating inclusion-wise minimal subsets $X \subseteq U$, called a \emph{minimal transversal} of $\mathcal H$, such that $X \cap e \neq \emptyset$ for every $e \in \mathcal E$.
The problem is one of the most challenging problems in \revised{the field of enumeration algorithms}, and the best-known algorithm is due to Fredman and Khachiyan~\cite{Fredman:complexity:1996}, which runs in total time $(|U| + N)^{O(\log (|U| + N))}$, where $N$ is the number of minimal transversals of $\mathcal H$. 
The existence of an \emph{output-polynomial time} algorithm, that is, it runs in total time $(U + |N|)^{O(1)}$, still remains open. 
Several papers~\cite{Ihler:complexity:1991,Klein:nearly:1995} pointed out some relation between the \emph{minimum} group Steiner tree problem and the \emph{minimum} \revised{transversal} problem to prove the hardness of approximation.
This relation also holds in the context of enumeration.

\begin{theorem}
    If there is an algorithm that solves {\sc Group Steiner Tree Enumeration} in output-polynomial time,
    then {\sc Minimal Transversal Enumeration} can be solved in output-polynomial time.
\end{theorem}
    
\begin{proof}
    Let $\mathcal H = (U, \mathcal E)$ be an instance of {\sc Minimal Transversal Enumeration}.
    Then, we construct a star graph $G$ as follows.
    The center of $G$ is denoted by $r$ and $G$ has a leaf vertex $\ell_{u}$ for each $u \in U$.
    For each $e \in \mathcal E$, we let $W_e = \{\ell_u : u \in e\}$ be a terminal set for $e$.
    It is not hard to see that $X$ is a minimal transversal of $\mathcal H$ if and only if $G[X \cup \{r\}]$ is a minimal solution of $(G, \{W_e : e \in \mathcal E\})$ for {\sc Group Steiner Tree Enumeration}.
    This indicates that {\sc Group Steiner Tree Enumeration} is at least as hard as {\sc Minimal Transversal Enumeration}. 
\end{proof}

\section{Minimal induced Steiner subgraphs for claw-free graphs}\label{sec:induced}
Another variant of {\sc Steiner Tree Enumeration} is {\sc Induced Steiner Tree Enumeration}, where the goal is to enumerate all inclusion-wise minimal subsets of vertices that induce Steiner subgraphs of given $(G, W)$.
Since every induced subgraph is defined as a subset of vertices, we may not distinguish them unless confusion arises.
Recall that a graph is \emph{claw-free} if it has no induced subgraph isomorphic to $K_{1,3}$, i.e., a star with three leaves.
Before describing the details of our proposed algorithm, we first observe that
{\sc Induced Steiner Tree Enumeration} on claw-free graphs is a generalization of {\sc Steiner Tree Enumeration}.

Let $G = (V, E)$ be a graph and let $W \subseteq V$ be terminals.
We begin with the line graph $L(G)$ of $G$ with $V(L(G)) = \set{v_e : e \in E}$ 
\revised{and two vertices $v_e$ and $v_e$ are adjacent in $L(G)$ if and only if they have a common end vertex}, 
and then construct a graph $H$ \revised{by adding vertices and edges to $L(G)$} as follows.
\revised{Starting from $H = L(G)$}, we add a vertex $w'$ to $H$ for each $w \in W$, and add an edge between $v_e$ and $w'$ for each $e \in \Gamma_G(w)$ in $H$.
Define $W_H = \set{v' : v \in W}$.
Then, the following theorem holds.

\begin{theorem}
    Let $T$ be a connected subgraph of $G$ and let $V_T = \{v_e : e \in E(T)\}$. Then, $T$ is a connected Steiner subgraph of $(G, W)$ if and only if $H[V_T \cup W_H]$ is a connected induced Steiner subgraph of $(H, W_H)$. 
\end{theorem}
\begin{proof}
    Suppose that $H[V_T \cup W_H]$ is a connected induced Steiner subgraph of $(H, W_H)$. Observe that $H[V_T]$ is connected in $H$. To see this, suppose that $H[V_T]$ is not connected. Since $H[V_T \cup W_H]$ is connected, there is a terminal $w \in W_H$ and components $X$ and $Y$ in $H[V_T]$ such that both $N_H(w) \cap X$ and $N_H(w)$ are nonempty. However, $N_H(w)$ induces a clique in $H$ for every $w \in W_H$, which is a contradiction. Thus, $H[V_T]$ is connected. Since $V_T$ contains at least one vertex in $N_H(w)$ for each $w \in W$ and $H[V_T]$ is connected, $T$ is a connected Steiner subgraph of $(G, W)$.

    Conversely, suppose that $T$ is a connected Steiner subgraph of $(G, W)$. For every pair of terminals $w$ and $w'$ in $W$, the edges of a path between them also induce a path between $w$ and $w'$ in $W_H$.
    This implies that $H[V_T \cup W_H]$ is a connected induced Steiner subgraph of $(H, W_H)$. 
\end{proof}

Since every line graph is claw-free, we conclude that {\sc Induced Steiner Tree Enumeration} on claw-free graphs is a generalization of {\sc Steiner Tree Enumeration}.

\begin{toappendix}

Our proposed algorithm is based on the \emph{supergraph technique}~\cite{DBLP:journals/jcss/CohenKS08,DBLP:conf/stoc/ConteU19,DBLP:journals/corr/abs-2004-09885,DBLP:journals/algorithmica/KhachiyanBBEGM08,DBLP:journals/dam/SchwikowskiS02}. 
Let us briefly describe the idea of this technique.
Let $G = (V, E)$ be a claw-free graph and $W \subseteq V$.
Let $\mathcal S \subseteq 2^V$ be the set of minimal induced Steiner subgraphs of $(G, W)$.
In this technique, we consider a directed graph $\mathcal G$ whose nodes correspond to the solutions $\mathcal S$, and whose arc set is defined so that $\mathcal G$ is strongly connected.
As $\mathcal G$ is strongly connected, we can enumerate all the solutions from an arbitrary one by traversing $\mathcal{G}$.
However, since strong connectivity is a ``global'' property and we do not know the entire node set $\mathcal S$ of $\mathcal G$, 
it would be nontrivial to ``locally'' define the set of arcs of $\mathcal G$, that is, define the neighborhood of each solution in $\mathcal G$. 
To this end, we define a ``distance'' measure $\sigma: \mathcal{S} \times \mathcal{S} \to \mathbb{Z}_{\ge 0}$ between two solutions.
If one can prove that 
\begin{itemize}
    \item[(1)] for $X, Y \in \mathcal S$, $\sigma(X, Y) = 0$ if and only if $X = Y$ and
    \item[(2)] for every pair of distinct solutions $X$ and $Y$, $X$ has a neighbor $Z$ in $\mathcal G$ with $\sigma(X, Y) > \sigma(Z, Y)$,
\end{itemize}
then $\mathcal G$ has a directed path from $X$ to $Y$, and hence $\mathcal G$ is strongly connected.
Specifically, for $X, Y \in \mathcal S$, we define $\sigma(X, Y) = |X \setminus Y|$.
As we will see later, the supergraph $\mathcal G$ defined by a simple construction does not satisfy condition (1).
To solve this issue, the second condition is relaxed to
\begin{itemize}
    \item[(2')] for every pair of distinct solutions $X$ and $Y$, there is a directed path from $X$ to $Z$ in $\mathcal G$ with $\sigma(X, Y) > \sigma(Z, Y)$,
\end{itemize}
which is sufficient to prove the strong connectivity of $\mathcal G$. 
We can easily see the following proposition from the definition of $\sigma$ and the minimality of solutions. 

\begin{proposition}\label{lem:ist:equal}
    Let $X$ and $Y$ be minimal induced Steiner subgraphs of $(G, W)$.
    Then, $\sigma(X, Y) = 0$  if and only if $X = Y$. 
\end{proposition}

To complete the description of our enumeration algorithm, we need to define the neighborhood relation in $\mathcal G$.
Given a connected vertex set $X \subseteq V$ \revised{with} $\revised{W} \subseteq X$, 
$\mu$ is a procedure that computes an arbitrary minimal induced Steiner subgraph $\comp[\revised{W}]{X}$ of $(G, W)$ that is contained in $X$. 
\revised{
Such a procedure is defined by a simple greedy algorithm that repeatedly removes a vertex $v$ from $X \setminus W$ as long as $G[X \setminus \{v\}]$ is an induced Steiner subgraph of $(G, W)$.
}
\revised{To define a neighbor of $X$ in $\mathcal G$, we need some observations.}
For each $v \in X \setminus W$, $G[X \setminus \{v\}]$ has two or more connected components \revised{as otherwise $G[X \setminus \{v\}]$ is an induced Steiner subgraph of $(G, W)$, which contradicts the minimality of $X$.}
Since $G$ is claw-free, there are exactly two connected components $C_1$ and $C_2$ in $G[X \setminus\set{v}]$.
\revised{This follows from the fact that if $G[X \setminus \set{v}]$ has three components, $v$ and three neighbors from these three components induce a claw $K_{1,3}$.}
As $X$ is a minimal induced Steiner subgraph of $(G, W)$, both components contain at least one terminal.

For each $w \in N(C_1) \setminus \{v\}$, we let $C^w_1 = \comp[(W \cap C_1)\cup \{w\}]{C_1 \cup \{w\}}$ and $C^w_2 = \comp[W \cap C_2]{C_2}$.
Note that $C^w_1$ is well-defined as $G[C_1 \cup \{w\}]$ is connected.
Let $P$ be an arbitrary shortest path between $w$ and $C^w_2$ that avoids $N(C^w_1) \setminus \{w\}$.
If such $P$ exists,  $C^w_1 \cup C^w_2 \cup V(P)$ is an induced Steiner subgraph of $(G, W)$ since $C^w_1 \cup C^w_2$ contains all the terminals and $P$ is an $N(C^w_1)$-$N(C^w_2)$ path.
Now, a neighbor $Z$ of $X$ is defined to be $\comp[W]{C^w_1 \cup C^w_2 \cup V(P)}$.
As all the vertices in $V(P)$ are cut vertices in $G[C^w_1 \cup C^w_2 \cup V(P)]$, we have $V(P) \subseteq Z$.
By the construction, $Z$ does not contain $v$ but does contain $w$.
Given this, we say that $Z$ is the \emph{neighbor of $X$ with respect to $(v, w)$} and define the arc set of $\mathcal G$ with this neighborhood relation.
We only define such a neighbor $Z$ when it is well-defined.
Clearly, every minimal induced Steiner subgraph of $(G, W)$ has $O(n^2)$ neighbors, which can be enumerated in \revised{$\order{n^2(n + m)}$} time.

Next, we show that for every pair of solutions $X$ and $Y$, there is a solution $Z$ reachable from $X$ that is ``closer to $Y$'' than $X$. 
The following lemma directly implies the strong connectivity of $\mathcal G$.

\begin{lemmarep}\label{lem:sc:iss}
    Let $X$ and $Y$ be distinct minimal induced Steiner subgraphs of $(G, W)$.
    Then, $\mathcal G$ has a directed path from $X$ to a minimal induced Steiner subgraph $Z$ of $(G, W)$ with $\sigma(X, Y) > \sigma(Z, Y)$.
\end{lemmarep}
\begin{proof}
    Let $v$ be a non-terminal vertex in $X \setminus Y$, and 
    $C_1$ and $C_2$ be the connected components in $G[X \setminus \{v\}]$.
    Since both components contain at least one terminal each, there is a path $P = (w_1, \ldots, w_k)$ between $N(C_1)$ and $N(C_2)$ such that $V(P) \subseteq Y$.
    We can assume that $w_1 \in N(C_1) \setminus \{v\}$, $w_k \in N(C_2) \setminus \{v\}$, and all the other vertices are not in $N(C_1)$ by appropriately choosing $P$.
    Let $C^{1}_1 = \comp[(W \cap C_1) \cup \set{w_1}]{C_1 \cup \{w_1\}}$ and $C^{1}_2 = \comp[W \cap C_2]{C_2}$.
    Since subpath $(w_2, \ldots, w_k)$ has no vertices in $N(C^1_1) \setminus \{w_1\}$, there is at least one shortest path $P_1$ between $w_1$ and $N(C^1_2)$ that avoids $N(C^1_1) \setminus \{w_1\}$.
    Define $X_1 = \comp[W]{C^1_1 \cup C^1_2 \cup V(P_1)}$ is the neighbor of $X$ with respect to $(v, w_1)$.
    If $P_1 = P$, then 
    \begin{linenomath}
    \begin{align*}
        \sigma(X_1, Y) &= |X_1 \setminus Y|\\
        &\le |(C^1_1 \cup C^1_2 \cup V(P_1)) \setminus Y|\\
        &\le |(C_1 \cup C_2) \setminus Y|\\
        &< |X \setminus Y|,
    \end{align*}
    \end{linenomath}
    and hence we are done.
    
    Suppose otherwise, that is, $P_1 \neq P$. 
    Recall that $V(P_1) \subseteq X_1$.
    $P_1$ contains $\{w_1, \ldots, w_{i - 1}\}$ and does not contain $w_i$ for some $1 < i \le k$.
    Let $v'$ be the vertex adjacent to $w_{i - 1}$ with $v \neq w_{i-2}$ in $P$.
    Let $C'_1$ and $C'_2$ be the connected components in $G[X_1 \setminus \{v'\}]$.
    Define $C^2_1 = \comp[(W \cap C'_1) \cup \{w_i\}]{C'_1 \cup \{w_i\}}$ and $C^2_2 = \comp[W \cap C'_2]{C'_2}$.
    We show that the following claim.
    
    \medskip
    \begin{claim}
        $C^2_2 \subseteq C_2$.
    \end{claim}
    
    \begin{claimproof}
        Let $P_1 = (w_1, \ldots, w_{i-1}, u_1, \ldots, u_t)$ with $u_1 = v'$ and $u_t \in N(C^1_2)$.
        Observe that $C'_2 = C^1_2 \cup \{u_2, \ldots u_t\}$.
        In the following, we prove that the vertices $u_2, \ldots, u_t$ vanish using the function $\mu$ regardless of its implementation, which proves the claim as $C^1_2 \subseteq C_2$.
        
        Since $P_1$ is a shortest path between $w_1$ and $N(C^1_2)$, every vertex in $\{u_2, \ldots, u_{t-1}\}$ is not adjacent to a vertex in $C^1_2$.
        This implies that those vertices are not contained in $C^2_2 = \comp[W \cap C'_2]{C'_2}$.
        Suppose that there is a minimal Steiner subgraph $C^2_2$ of $(G[C'_2], W \cap C'_2)$ containing $u_t$.
        We assume that $C'_2$ contains at least two terminals as otherwise $C^2_2$ consists of exactly one vertex, which is terminal.
        Since $u_t$ is in $C^2_2$, $C^2_2$ contains an induced path between two terminals in $C^2_2$ passing through $u_t$.
        Then, the two adjacent vertices of $u_t$ in this path together with $u_t$ and $u_{t - 1}$ form an induced claw, which contradicts the fact that $G$ is claw-free.
    \end{claimproof}
    
    By the same argument as above, the neighbor $X_2$ of $X_1$ with respect to $(v', w_i)$ is well-defined.
    We inductively compute the neighbor $X_i$ of $X_{i - 1}$ unless $P_i$ contains a vertex that does not belong to $V(P)$.
    Eventually, we have $X_i$ that consists of $C^i_1 \cup C^i_2 \cup V(P_i)$, where $C^i_1 \subseteq C_1 \cup V(P)$, $C^i_2 \subseteq C_2$, and $V(P_i) \subseteq V(P)$.
    Then, we have 
    \begin{linenomath}
     \begin{align*}
        \sigma(X_i, Y) &= |X_i \setminus Y|\\
        &\le |(C_1 \cup C_2 \cup V(P)) \setminus Y|\\
        &\le |(C_1 \cup C_2) \setminus Y|\\
        &< |X \setminus Y|,
    \end{align*}
    \end{linenomath}
    which completes the proof of lemma. 
\end{proof}

By \Cref{lem:sc:iss}, $\mathcal G$ is strongly connected. 
Thus, by using a standard graph search algorithm on $\mathcal G$, we can enumerate all the minimal induced Steiner subgraphs of $(G, W)$ in \revised{$\order{n^2(n + m)}$ delay}. 
\end{toappendix}

\begin{theorem}
    \label{thm:isse}
    \textsc{Induced Steiner Subgraph Enumeration} can be solved in \revised{$\order{n^2(n + m)}$} delay using exponential space on claw-free graphs.
\end{theorem}

\section*{Acknowledgments} 
This work is partially supported by JSPS Kakenhi Grant Numbers
JP19K20350, 
JP20K19742, 
JP20H05793, 
JP21K17812, 
JP21H05861, 
JP21H03499, 
JST CREST Grant Number JPMJCR18K3, 
and JST ACT-X Grant Number JPMJAX2106, Japan 




{
    \bibliographystyle{plain}
    \bibliography{kurita}
}









\end{document}

%% file: macro.tex
\newcommand{\set}[1]{\{#1\}}

\newcommand{\size}[1]{\left|#1\right|}
\newcommand{\order}[1]{O\mathopen{}\left(#1\right)\mathclose{}}

\newcommand{\NP}{{NP}\xspace}

\crefname{rrule}{Reduction rule}{Reduction rules}
\newenvironment{claim}[1]{\par\noindent\underline{Claim:}\space#1}{}
\newenvironment{claimproof}[1]{\par\noindent\underline{Proof of Claim:}\space#1}{\hfill $\square$}

\newcommand{\etal}{\textit{et~al.}\xspace}

\makeatletter
\@ifpackageloaded{algorithm2e}{
\SetKwProg{Fn}{Function}{}{}
\SetKwProg{Procedure}{Procedure}{}{}
\SetKwProg{Subprocedure}{Subprocedure}{}{}
\SetKwComment{tcc}{//}{}
\SetKwFunction{Output}{Output}%
\SetKwFunction{EnumSTP}{E-STP}
\SetKwFunction{EnumMST}{E-MST}
\SetKwFunction{EnumSol}{E-Sol}
\SetKwFunction{FindSTP}{F-STP}
\SetKwFunction{RecMST}{Rec-MST}
\SetKwFunction{FindMSF}{Find-MSF}
\SetKwFunction{checkMST}{MST-exists}
\SetKwFunction{checkSol}{Sol-exists}
\SetKwFunction{extendible}{extendible}
\SetKwFunction{reduce}{reduce}
\SetKw{Continue}{continue} 
\SetKwInput{AlgInput}{Input}
\SetKwInput{AlgOutput}{Output}
\SetKwInOut{AlgPrecondition}{Pre-conditions}
\SetKwInOut{AlgInvariant}{Invariants}
}{}
\makeatother

\makeatletter

\newcolumntype{\expand}{}
\long\@namedef{NC@rewrite@\string\expand}{\expandafter\NC@find}

\newcommand{\comp}[2][]{\ifthenelse{\equal{#1}{}}{\mu(#2)}{\mu(#2, #1)}}